\documentclass[aps,pre,twocolumn,superscriptaddress,showpacs]{revtex4-1}
\usepackage{graphicx}
\usepackage{dcolumn}
\usepackage{bm}
\usepackage{epsfig}
\usepackage{color}
\usepackage{soul}

\begin{document}


\title{Enhanced flow rate by the concentration mechanism of Tetris
  particles when discharged from a hopper with an obstacle}


\author{Guo-Jie Jason Gao}
\email{gao@shizuoka.ac.jp}
\affiliation{Department of Mathematical and Systems Engineering,
  Shizuoka University, Hamamatsu, Shizuoka 432-8561, Japan}

\author{Fu-Ling Yang} \affiliation{Department of Mechanical
  Engineering, National Taiwan University, Taipei 10617, Taiwan}

\author{Michael C. Holcomb} \affiliation{Department of Physics and
  Geosciences, Angelo State University, San Angelo, TX 76909-0904,
  USA}

\author{Jerzy Blawzdziewicz} \affiliation{Department of Physics, Texas
  Tech University, Lubbock, TX 79409-1051, USA}
\affiliation{Department of Mechanical Engineering, Texas Tech
  University, Lubbock, TX 79409-1021, USA}

\date{\today}

\begin{abstract}
We apply a holistic 2D Tetris-like model, where particles move based
on prescribed rules, to investigate the flow rate enhancement from a
hopper. This phenomenon was originally reported in the literature as a
feature of placing an obstacle at an optimal location near the exit of
a hopper discharging athermal granular particles under gravity. We
find that this phenomenon is limited to a system of sufficiently many
particles. In addition to the waiting room effect, another mechanism
able to explain and create the flow rate enhancement is the
concentration mechanism of particles on their way to reaching the
hopper exit after passing the obstacle. We elucidate the concentration
mechanism by decomposing the flow rate into its constituent variables:
the local area packing fraction $\phi_l^E$ and the averaged particle
velocity $v_y^E$ at the hopper exit. In comparison to the case without
an obstacle, our results show that an optimally placed obstacle can
create a net flow rate enhancement of relatively weakly driven
particles, caused by the exit-bottleneck coupling if $\phi_l^E >
\phi_o^c$, where $\phi_o^c$ is a characteristic area packing fraction
marking a transition from fast to slow flow regimes of Tetris
particles. Utilizing the concentration mechanism by artificially
guiding particles into the central sparse space under the obstacle or
narrowing the hopper exit angle under the obstacle, we can create a
man-made flow rate peak of relatively strongly-driven particles that
initially exhibit no flow rate peak.  Additionally, the enhanced flow
rate can be maximized by an optimal obstacle shape, particle
acceleration rate towards the hopper exit, or exit geometry of the
hopper.
\end{abstract}


\maketitle


\section{Introduction}
\label{introduction}
Both experimentally and numerically, placing an obstacle at an optimal
distance away from the exit of a hopper has been shown to enhance the
gravity-driven granular hopper flow rate on the order of ten percent
\cite{zuriguel11, zuriguel14, lozano12, alonso-marroquin12,
  zuriguel15, alonso-marroquin16}. This strategy has been shown to be
effective not only on passive granular particles but also on
self-governing species
\cite{helbing00,rosa03,helbing05,zuriguel15_1,zuriguel16}. One of the
possible explanations for the enhanced flow rate is the waiting room
effect, wherein particles are slowed down by the obstacle and then
accelerate within the void underneath it on their way towards the
hopper exit \cite{rosa03,zuriguel11,lozano12,alonso-marroquin12,
  alonso-marroquin16}. However, some studies are either unable to
reproduce this phenomenon \cite{katsuragi17,katsuragi18}, or able to
reproduce it even under conditions in which the void space below the
obstacle has been eliminated through the introduction of a special
obstacle shape that diminishes the waiting room effect
\cite{alonso-marroquin16}. Conventional experiments and numerical
approaches, governed by Newtonian dynamics, contain multiple competing
mechanisms such as interparticle collaborative motion and particle
acceleration due to gravity; these cannot be easily decomposed and
inspected separately. To isolate the roles of these competing
mechanisms, a more primitive dynamic model is needed.

In our previous studies, we showed that the interparticle friction,
particle dispersity, and obstacle geometry are not directly
responsible for the enhanced flow rate \cite{gao19}. We then proposed
a 2D Tetris-like model, where particles move according to prescribed
rules rather than in response to forces in order to switch off
interparticle collaborative motion. Using our model, we still observed
the enhanced flow rate; therefore, the collaborative motion of
particles via Newtonian dynamics is also not the key mechanism
\cite{gao18, gao19}. Another more simplified Tetris-like model, where
particles can only move diagonally, was proposed to study the packing
behavior of granular materials under vibration \cite{nicodemi97}. In
contrast to reductionist models, such as discrete element methods that
preserve enough details to quantitatively reproduce an aimed physical
phenomenon, Tetris-like models are holistic and focus on similarities
between different nonequilibrium systems containing animate or
inanimate discrete particles. The results of our model suggest that
the concentration of particles arriving at the hopper exit is
essential to the observed flow rate peak, as the local area packing
fraction of particles $\phi_l$ (defined in the cited reference
\cite{gao19} and again in Sec. \ref{peaking_mechanism}) increases near
the hopper exit and can become larger than its value near the
obstacle. In contrast, we do not observe the same pattern of $\phi_l$
variation in a system that exhibits no flow rate peak.

Following this finding, this work further uses the 2D Tetris-like
model to explore how the flow rate is influenced by the number of
particles in the hopper; the concentration behavior of particles below
the obstacle; and the waiting room effect, as it relates to obstacle
geometry, particle acceleration, and exit geometry of the hopper. We
find that the flow rate peaking phenomenon is limited to a system of
sufficiently many particles, and that an obstacle in a hopper
containing too few particles only reduces its flow rate. In addition,
placing an obstacle in a hopper of sufficiently many particles can
reduce the particle area packing fraction more than without an
obstacle and, therefore, particles move faster. The net effect is two
downstream particle flows concentrating at the hopper exit with an
area packing fraction able to be greater than a characteristic value,
which triggers an exit-bottleneck coupling. This mechanism is directly
responsible for the flow rate peaking phenomenon if particles are
relatively weakly-driven and the obstacle is optimally
placed. Moreover, we utilize the concentration mechanism to prove that
a flow rate enhancement can be generated in a system discharging
relatively strongly-driven particles that originally exhibits no flow
rate peak. This can be accomplished by artificially increasing the
horizontal driving strength asymmetrically of the particle flowing
down stream of the obstacle, or by narrowing the hopper exit angle
under the obstacle.  Finally, we evaluate the waiting room effect and
show the existence of an optimal obstacle geometry, particle
acceleration rate towards the hopper exit, and exit geometry can
maximize or create an enhanced flow rate.

Below we elaborate on our Tetris-like model which generates the
probability-driven hopper flow in section \ref{tetris model}, followed
by quantitative investigations of the hopper flow rates under
different conditions in section \ref{results_and_discussions} with
discussions. We conclude our study in section \ref{conclusions}.

\section{The Tetris-like model}
\label{tetris model}
\begin{figure}
\includegraphics[width=0.40\textwidth]{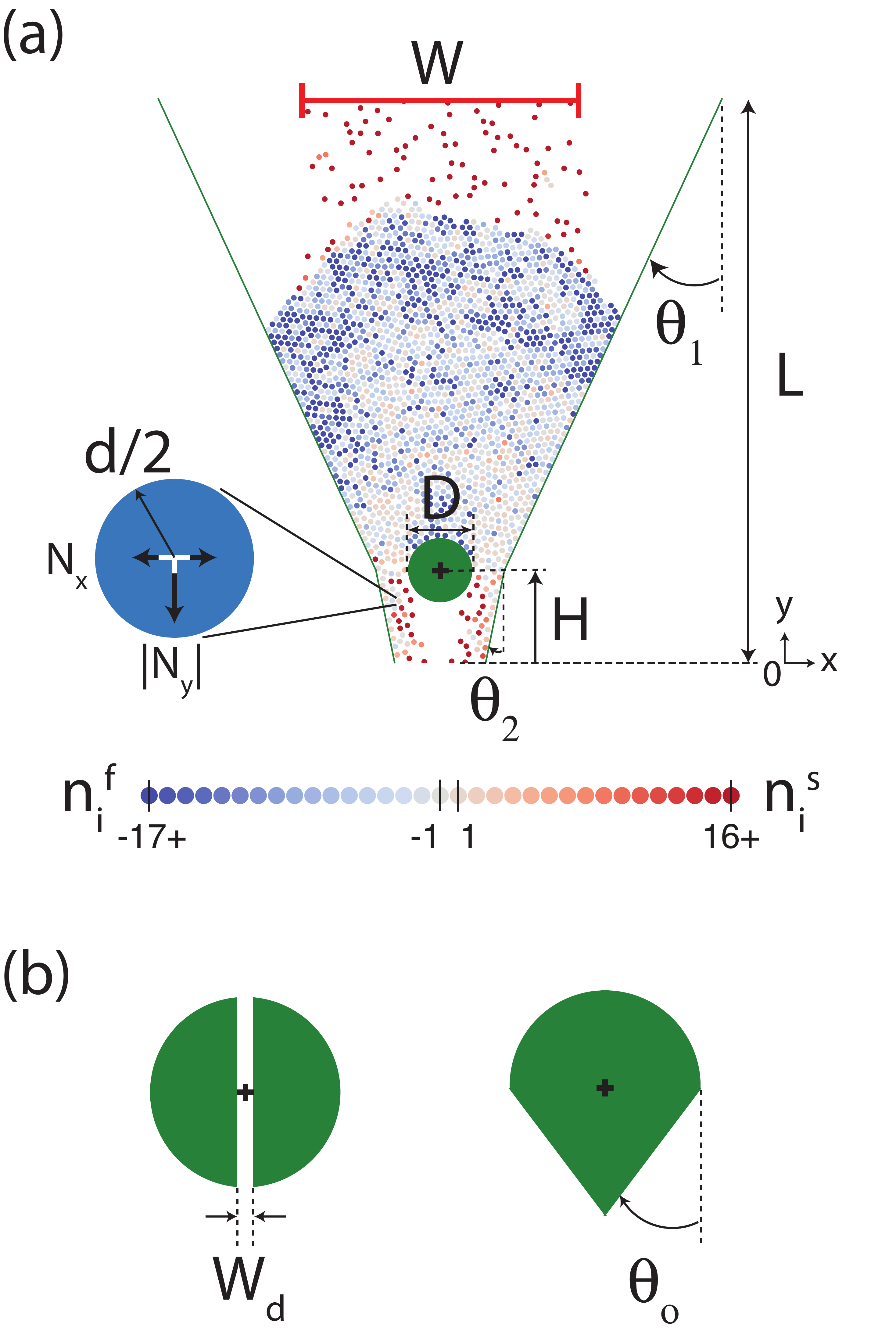}
\caption{\label{fig:tetris_model} (Color online) (a) A snapshot of the
  system setup. A symmetric hopper (green lines) with equal height and
  top-width $L$, a hopper angle $\theta_1$, and an exit angle
  $\theta_2$ contains a round obstacle (green circle) of diameter $D$,
  placed along its centerline at a height H above its exit. Particles
  discharged from the hopper reenter it from its top boundary with
  their $x$ positions randomized in between $W = 0.5L$.  The discrete
  red-blue colors represent the value of $n_i^s>0$ or $n_i^f<0$,
  recording the position-update history of consecutive successes or
  failures. (b) The other two kinds of obstacles, a hollow circle and
  a semicircle-triangle, used in the study. Their circular parts have
  the same diameter $D$, and they are also located at a height $H$
  above the hopper exit, measured from their centers (cross marks).}
\end{figure}

To study the simplest 2D granular hopper flow without invoking
Newtonian dynamics, we propose a model, named after the video game
Tetris, where objects fall one at a time following some prescribed
rules within a confined space. Per position-update cycle in our model,
each particle $i$ of uniform diameter $d$ attempts to move exactly
once from its current $x$ and $y$ positions $(x_i^{old}, y_i^{old})$
to $(x_i^{new}, y_i^{new})$, specified by
\begin{equation} \label{tetris_algorithm}
\begin{array}{l}
x_i^{new} = x_i^{old} + {N_x}(0,{\alpha _x}\sigma )\\ y_i^{new} =
y_i^{old} - \left| {{N_y}(0,{\alpha _y}\sigma )}
\right|{r_s}^{n_i^{s}}
\end{array},
\end{equation}
where $N_x$ and $N_y$ are normal distribution functions having zero
means and standard deviations ${\alpha_x}\sigma$ and
${\alpha_y}\sigma$, and specify displacements of particle $i$. The two
independent control parameters ${\alpha_x}$ and ${\alpha_y}$ determine
the driving strengths in the horizontal $x$ and vertical $y$
directions, respectively. The absolute value about $N_y$ guarantees
that a particle always moves in only one direction towards the hopper
exit. We chose a moderate $\sigma=0.05d$ so that on average particles
exit the hopper containing no obstacle with their trajectories
parallel to the hopper walls \cite{gao18}. A move attempt is only
realized if it creates no overlap between any object in the
system. Otherwise, the move attempt is rejected and the attempted
particle stays still. Which particle moves first is determined by a
random sequence, regenerated at the beginning of each position-update
cycle. Additionally, each particle $i$ remembers its position-update
history, recorded as a monotonically increasing number $n_i^{s}>0$ or
decreasing number $n_i^{f}<0$ for consecutive successful or failed
attempts. Whenever one parameter becomes nonzero, the other is reset
to zero. The speed-up rate $r_s \ge 1$ in the vertical $y$ direction
mimics the effect of particle acceleration due to gravity during free
fall. A particle that successfully updates its position $n_i^s$ times
can attempt a longer jump due to acceleration by a factor of
${r_s}^{n_i^{s}}$ during the next position-update cycle. We do not
include $n_i^f$ in Eqn. \ref{tetris_algorithm} because it is
intrinsically related to particle rebounding behavior, which is not
allowed in the current model. The Tetris-like model can also be viewed
as a 2D cellular automaton in a sense that the size of a cell,
surrounded by a circular excluding zone of diameter $d$, is
infinitesimally small down to the machine precision. Additional
details about the Tetris-like model can be found in our previous
studies \cite{gao18, gao19}.

The geometrically symmetric hopper, measuring $L=83d$ in height with a
fixed hopper angle $\theta_1=0.4325$ (rad) and a changeable exit angle
$\theta_2$, contains $N$ randomly placed particles at the beginning of
each simulation. The the hopper's orifice size is a function of
$\theta_1$ and $\theta_2$. In all figures except
Fig. \ref{fig:kink_dependence_py0143} and
Fig. \ref{fig:kink_dependence_py0180}, $\theta_2=\theta_1$, which
gives an orifice width of about $6.366d$. In
Fig. \ref{fig:kink_dependence_py0143} and
Fig. \ref{fig:kink_dependence_py0180}, the orifice size decreases with
increasing $\theta_2$, where the maximum value of $\theta_2=0.5$
corresponds to the minimum orifice size of about $4.061d$. To conserve
the total number of particles within the system, $N$, a particle
coming out of the hopper from its exit will reenter it from above with
the particle's new $x$ position randomized in-between $W=[-L/4,L/4]$,
as shown by a snapshot in Fig. \ref{fig:tetris_model}(a). After a
simulation reaches its steady state, we measure the flow rate $J_o$ in
terms of the average number of particles passing the orifice of the
hopper containing no obstacle per position-update cycle. Similarly, we
measure $J_a$ when the hopper contains a circular obstacle or an
obstacle with a semicircular component of diameter $D=0.112L$ whose
center is placed along the symmetric axis of the hopper at a height
$H$ above its exit. We use obstacles of different shapes to study the
effect of the obstacle geometry on the flow rate, such as an obstacle
with a hollow duct of width $W_d$ or another one composed of a
semicircle top and isosceles triangle bottom specified by an angle
$\theta_o$ to the vertical, as depicted in
Fig. \ref{fig:tetris_model}(b). Each data point of $J_o$ or $J_a$ is
obtained using $45$ different initial conditions in steady state
followed by 990,000 position-update cycles to ensure adequate sampling
and to obtain error bars representing the standard deviation of each
measured quantity.

\section{Results and Discussions}
\label{results_and_discussions}
Below, we first plot the normalized hopper flow rates $J_a/J_o$ as a
function of the total number of particles in the system $N$. We then
show the flow rate enhancement is caused by the particle concentration
mechanism below the obstacle by measuring the constituent variables,
the local area packing fraction $\phi_l$ and the averaged particle
velocity $v_y$ along the $y$ direction, of $J_a/J_o$. After that we
investigate the importance of the particle concentration mechanism
within the void below the obstacle in-between $y=[0,H]$ by
artificially making the horizontal driving strength $\alpha_x$
anisotropic in this region, which mobilizes particles in the void
space and increases the local area packing fraction $\phi_l$ near the
hopper exit. We show that this strategy is actually a way of utilizing
the concentration mechanism by manipulating particles below the
obstacle. We also use an obstacle with a hollow duct of variable width
$W_d$ to examine the sensitivity of the particle concentration
mechanism. Finally, we explore the waiting room effect on the flow
rate enhancement by using a semicircular obstacle with a triangular
half of adjustable $\theta_o$, changing the vertical speed-up rate
$r_s$, or varying the exit angle of the hopper $\theta_2$.

\subsection{The effect of the total number of particles in the system}
\label{Effect_N_dependence_J}
\begin{figure}
\includegraphics[width=0.40\textwidth]{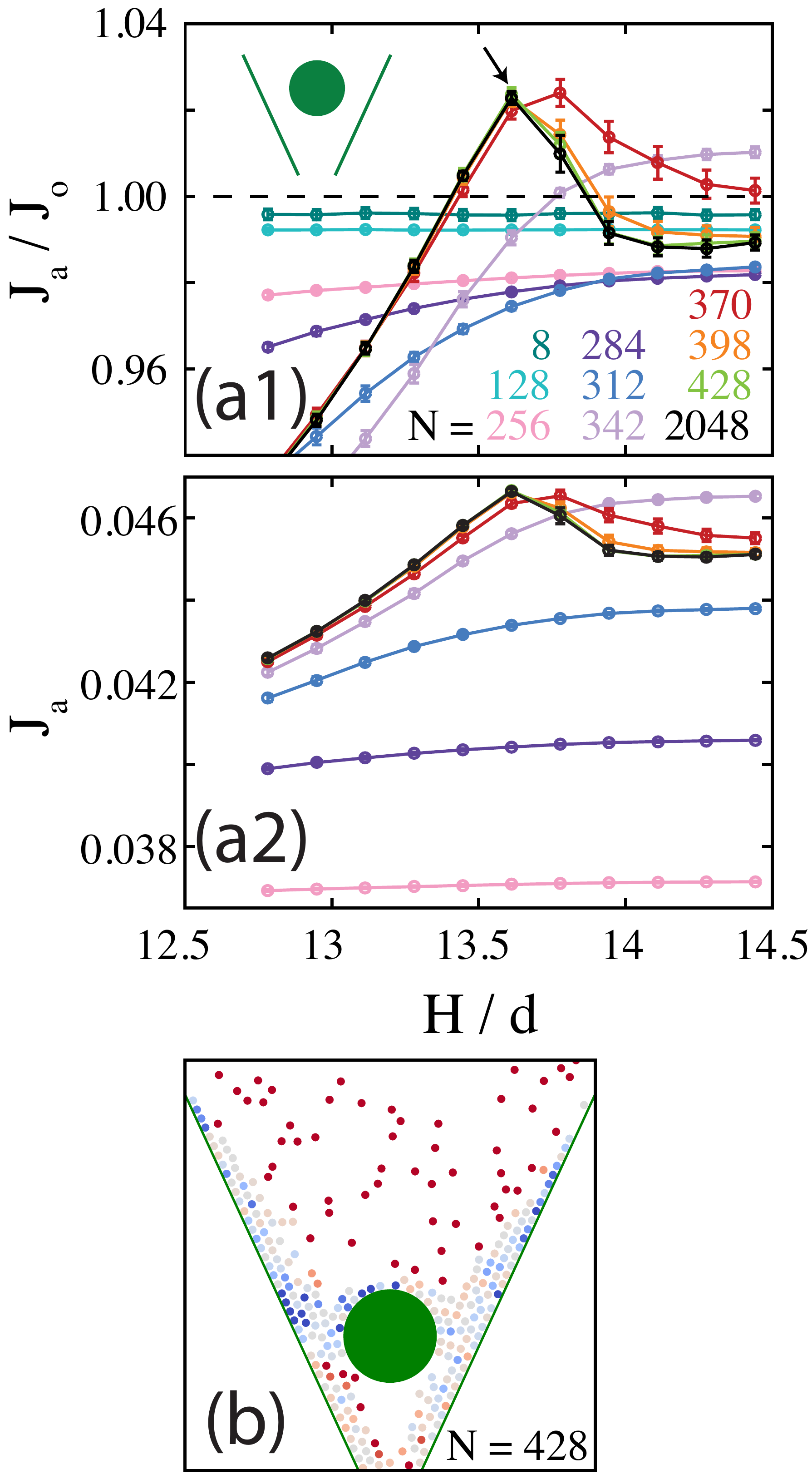}
\caption{\label{fig:N_dependence_J} (Color online) (a1) Normalized
  hopper flow rates $J_a/J_o$ measured at the exit of a hopper with
  $\theta_2=\theta_1$ and containing a round obstacle. The total
  number of particles in the system $N=8$ (dark green), $128$ (cyan),
  $256$ (pink), $284$ (dark purple), $312$ (navy), $342$ (purple),
  $370$ (red), $398$ (orange), $428$ (green), and $2048$ (black). The
  driving strengths $\alpha_x=1.0$, $\alpha_y=0.333$, and the speed-up
  rate $r_s=1.0$. (a2) Same plot as (a1) except unnormalized hopper
  flow rates $J_a$ are shown. The curves of $N=8$ and $128$ are not
  included because their $J_a$ values are outside the visible vertical
  range. (b) A snapshot focusing on the lower half of the hopper with
  $N=428$ where an obstacle is placed as indicated by the arrow in
  (a). Particles are colored by the blue-red scheme in
  Fig. \ref{fig:tetris_model}(a).}
\end{figure}

To test the effect of the total number of particles in the hopper $N$
on the flow rate, we tried ten different system sizes between $N=8$
and $2048$ and measured the corresponding $J_a/J_o$. The results with
$\alpha_x=1.0$, $\alpha_y=0.333$, and $r_s=1.0$ are shown in
Fig. \ref{fig:N_dependence_J}(a1) for normalized $J_a/J_o$ and
Fig. \ref{fig:N_dependence_J}(a2) for unnormalized $J_a$,
respectively. Initially, $J_a$ increases rapidly with $N$.  When $N$
is smaller than about $300$, the obstacle only slows down the
normalized flow rate monotonically with increasing $N$ and diminishing
$H$. However, when $N \ge 342$, $J_a/J_o$ becomes greater than $1.0$,
reflecting that $H$ exceeds a characteristic value, and exhibits a
local peak. The enhanced flow rate can be achieved only when the
obstacle is placed within an $H$ range that shrinks with increasing
$N$ and finally saturates as $N$ approaches about $428$, the minimum
number of particles required to reproduce the results at the large
system size limit. A snapshot where a flow rate peak happens with
$N=428$ is shown in Fig. \ref{fig:N_dependence_J}(b). The outcome of
the system size dependence test suggests that the flow rate
enhancement is limited to a system of a sizable number of
particles. As we reveal in the following sections, as particles move
through the two channels between the obstacle and the hopper walls
their trajectories concentrate and a slightly sparser but faster
particle flow is formed and discharged at the hopper exit, as compared
to the case without the obstacle. This produces a larger than unity
$J_a/J_o$. A hopper containing too few particles cannot sustain the
associated concentration mechanism crucial for the flow rate peak.

\subsection{The concentration mechanism causing the flow rate peaking phenomenon in the Tetris-like model}
\label{peaking_mechanism}
To understand the flow rate peaking phenomenon, we decompose the flow
rate into ${J_a} = \phi _lv_y{w}$ in terms of the local area packing
fraction $\phi_l$, the width-averaged particle velocity $v_y$ (in the
$y$ direction), and the passage horizontal width $w$ at vertical
location $y_c$ where $v_y$ is measured. Specifically, we define
$\phi_l(y_c)$ at $y=y_c$ as the ratio of the total area of particles,
whose centers reside within the space confined by $y = {y_c} +
\Delta_y/2$, $y = {y_c} - \Delta_y/2$, and the two hopper walls, to
the available area of the confined space. Similarly, we define
${v_y}({y_c}) = \Delta_y/\bar t$ at $y=y_c$, where $\bar t$ is the
average position-update cycles required for a particle travels from $y
= {y_c} + \Delta_y/2$ to $y = {y_c} - \Delta_y/2$. We chose a constant
$\Delta_y = 0.03L \approx 2.49d$ throughout this study. For a given
system setup, we randomly choose $10$ trials out of the $45$ initial
conditions to obtain the means and error bars of $\phi_l$ and
$v_y$. Superscripted with a capital E, the calculated hopper flow
rates at the exit with or without the obstacle may be expressed as
${J_a^E} = \phi _l^Ev_y^E{w^E}$ or ${J_o^E} = \phi
_{lo}^Ev_{yo}^E{w^E}$, respectively. The local values of $\phi _l^E$,
$v_y^E$, $\phi _{lo}^E$, and $v_{yo}^E$ are measured at $y_c^E =
0.015L \approx 1.245d$, only slightly above the hopper exit at $y=0$
for an accurate estimation, and $w^E$ is the corresponding orifice
width at $y_c^E$. An extra subscript letter $o$ is added to
distinguish the values measured without an obstacle. Putting
everything together, we have the normalized flow rate ${{{J_a^E}}
  \mathord{\left/ {\vphantom {{{J_a^E}} {{J_o^E}}}} \right.
    \kern-\nulldelimiterspace} {{J_o^E}}} = {{(\phi _l^E}
  \mathord{\left/ {\vphantom {{(\phi _l^E} {\phi _{lo}^E)({{v_y^E}
            \mathord{\left/ {\vphantom {{v_y^E} {v_{yo}^E}}} \right.
              \kern-\nulldelimiterspace} {v_{yo}^E}}}}} \right.
    \kern-\nulldelimiterspace} {\phi _{lo}^E)({{v_y^E} \mathord{\left/
        {\vphantom {{v_y^E} {v_{yo}^E}}} \right.
        \kern-\nulldelimiterspace} {v_{yo}^E}}}})$ in terms of the
product of the normalized local area packing fraction ${{\phi _l^E}
  \mathord{\left/ {\vphantom {{\phi _l^E} {\phi _{lo}^E}}} \right.
    \kern-\nulldelimiterspace} {\phi _{lo}^E}}$ and the normalized
averaged particle velocity ${{v_y^E} \mathord{\left/ {\vphantom
      {{v_y^E} {v_{yo}^E}}} \right.  \kern-\nulldelimiterspace}
  {v_{yo}^E}}$ at the hopper exit.

\begin{figure}
\includegraphics[width=0.40\textwidth]{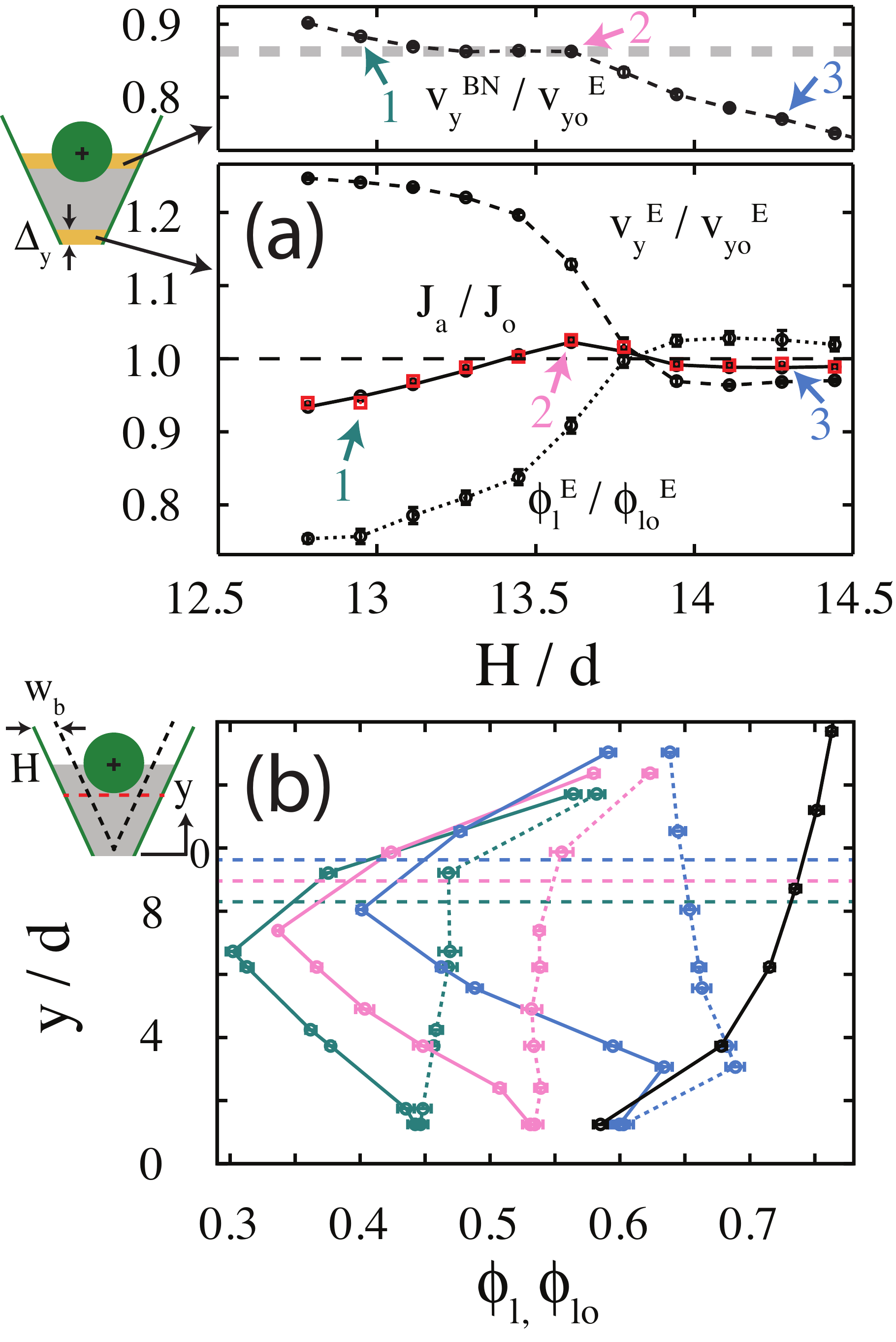}
\caption{\label{fig:factorized_flowrate1} (Color online) (a) The lower
  box shows a plot of normalized local area packing fraction $\phi
  _l^E/\phi _{lo}^E$ (dotted line), normalized averaged particle
  velocity $v_y^E/v_{yo}^E$ (dashed line), measured at $y_c = 0.015L
  \approx 1.245d$ (lower yellow region in the inset). The calculated
  normalized flow rate using the product ${J_a^E}/{J_o^E} = (\phi
  _l^E/\phi _{lo}^E)(v_y^E/v_{yo}^E)$ is also marked (red
  squares). The corresponding normalized hopper flow rate $J_a/J_o$
  (solid line) is copied from Fig. \ref{fig:N_dependence_J}(a1) for
  reference. The upper box shows a plot of $v_y^{BN}/v_{yo}^E$,
  measured at bottleneck $y_c = H - 0.015L$ (upper yellow region in
  the inset). The dashed line indicates its plateau value. The driving
  strengths $\alpha_x=1.0$, $\alpha_y=0.333$, the speed-up rate
  $r_s=1.0$, and $N=2048$. (b) Averaged local area packing fraction
  $\phi_l$ in the shaded zone in-between $y=[0,H]$ for the three
  selected cases labeled in (a), where the obstacle is placed at 1
  (dark green), 2 (pink), and 3 (navy). The dotted lines are $\phi_l$
  next to the hopper wall within a stripe of width $w_b \approx
  3.669d$, limited by the black dashed lines marked on the inset. The
  horizontal dashed lines indicate the lowest $y$ positions of the
  obstacle. The black line is the corresponding averaged local area
  packing fraction $\phi_{lo}$ while the hopper contains no obstacle.}
\end{figure}

In the lower box of Fig. \ref{fig:factorized_flowrate1}(a), we plot
$J_a/J_o$ with $\alpha_y=0.333$ and $N=2048$, copied from
Fig. \ref{fig:N_dependence_J}(a1), together with its corresponding
measured constituent variables ${{\phi _l^E} \mathord{\left/
    {\vphantom {{\phi _l^E} {\phi _{lo}^E}}} \right.
    \kern-\nulldelimiterspace} {\phi _{lo}^E}}$ and ${{v_y^E}
  \mathord{\left/ {\vphantom {{v_y^E} {v_{yo}^E}}} \right.
    \kern-\nulldelimiterspace} {v_{yo}^E}}$ that give the calculated
$J_a^E/J_o^E$. The calculated $J_a^E/J_o^E$ using $\phi _l^E/\phi
_{lo}^E$ and $v_y^E/v_{yo}^E$ are in close agreement to the measured
$J_a/J_o$, supporting our approach of variable decomposition and its
estimation. As the obstacle is placed at a higher place above the
hopper exit with increasing $H/d$, the two channels between the
obstacle and the hopper walls become wider and allow more particles to
move through. As a result, this creates a monotonically increasing
$\phi _l^E/\phi _{lo}^E$ at the hopper exit. Correspondingly, a higher
packing fraction comes with higher particle clogging probability,
which reduces particle velocity. Therefore, we simultaneously observe
a monotonically decreasing $v_y^E/v_{yo}^E$ at the hopper
exit. Placing the obstacle too close to the hopper exit blocks too
many particles, yielding in a low $\phi _l^E/\phi _{lo}^E$. On the
other hand, placing the obstacle too far from the hopper exit results
in $v_y^E/v_{yo}^E<1$ due to a high clogging probability from too many
particles being discharged out of the hopper. Only an optimally placed
obstacle blocks just enough particles so that $\phi _l^E/\phi _{lo}^E$
is slightly lower than $1$ while also allowing particles to be
discharged faster at the hopper exit with $v_y^E/v_{yo}^E>1$. This
creates an enhanced flow rate, where a characteristic transition
happens when $\phi_l^E>\phi_o^c \approx 0.55$, as discussed in the
appendix Sec. \ref{flow_property} (variation of $\phi_l^E$ is shown in
Fig. \ref{fig:factorized_flowrate1}(b) below). The characteristic
transition experiences a positive feedback from the shrinking geometry
of the hopper, which raises $\phi_l^E$ with decreasing $v_y^E$ while
particles move through the hopper exit. Reciprocally, a slower $v_y^E$
increases $\phi_l^E$, which helps maintain the condition of
$\phi_l^E>\phi_o^c$ until particles are discharged from the hopper. In
the upper box of Fig. \ref{fig:factorized_flowrate1}(a), we plot
normalized bottleneck velocity $v_y^{BN}/v_{yo}^E$, measured at $y_c =
H - 0.015L$, which exhibits a plateau before the flow rate peak and
decreases again afterwards. The plateau is a signal that upstream
particles at the bottleneck start to sense the downstream
characteristic transition.

In Fig. \ref{fig:factorized_flowrate1}(b), we plot the local area
packing fraction $\phi _l(y)$ under the obstacle in-between $y=[0,H]$
(see the shaded zone in the inset) of the three selected cases,
labeled in Fig. \ref{fig:factorized_flowrate1}(a). We chose the zone
in this way because $\phi _l$ exhibits the most significant variation
in this region, and its pattern of variation depends on whether a
system exhibits an enhanced flow rate or not, as reported in our
previous study \cite{gao19}. In all cases, $\phi _l$ first decreases
to a local minimum somewhere below the obstacle due to the available
void therein, after which $\phi _l$ increases because of the
concentration of particles on their way towards the hopper exit. In
the case after the flow rate peak (labeled as 3 in
Fig. \ref{fig:factorized_flowrate1}(a)), $\phi _l$ decreases again to
facilitate the discharging of extremely densely-packed particles with
$\phi _l>0.6$ near the hopper exit. To observe the true variation of
$\phi _l$ without the interruption of the void below the obstacle, we
also plot $\phi _l$ next to the hopper wall within a stripe of width
$w_b \approx 3.669d$, which is the narrowest width between the
obstacle and the hopper wall when the obstacle is placed at the lowest
position of the three selected cases. Both before and at the flow rate
peak, $\phi _l$ next to the hopper wall still exhibits some decrease
immediately after particles pass the narrowest channel segments, which
implies particle acceleration. Afterwards, $\phi _l$ stays
approximately constant until reaching the hopper exit. The fact that
$\phi _l$ can maintain a constant value is not trivial and is caused
by the shrinking geometry of the hopper exit generating a
concentrating effect, which goes away as soon as we reduce the exit
angle $\theta_2$, as shown later in
Fig. \ref{fig:kink_dependence_py0143}(b). In contrast, $\phi _{lo}$
exhibits a monotonic decrease below the obstacle. By comparing $\phi
_l$ with $\phi _{lo}$, we can see that introducing an obstacle into
the hopper changes the discharging mechanism. When the hopper contains
an obstacle near its exit, the two concentrated particle flows merge
near the hopper exit before discharge. When the hopper contains no
obstacle, on the other hand, a dense group of particles with
$\phi_{lo}>\phi_o^c$ above the hopper exit has to dismantle first
before exiting the hopper, as suggested by the progressively
decreasing $\phi _{lo}$ towards the hopper exit.

\begin{figure}
\includegraphics[width=0.267\textwidth]{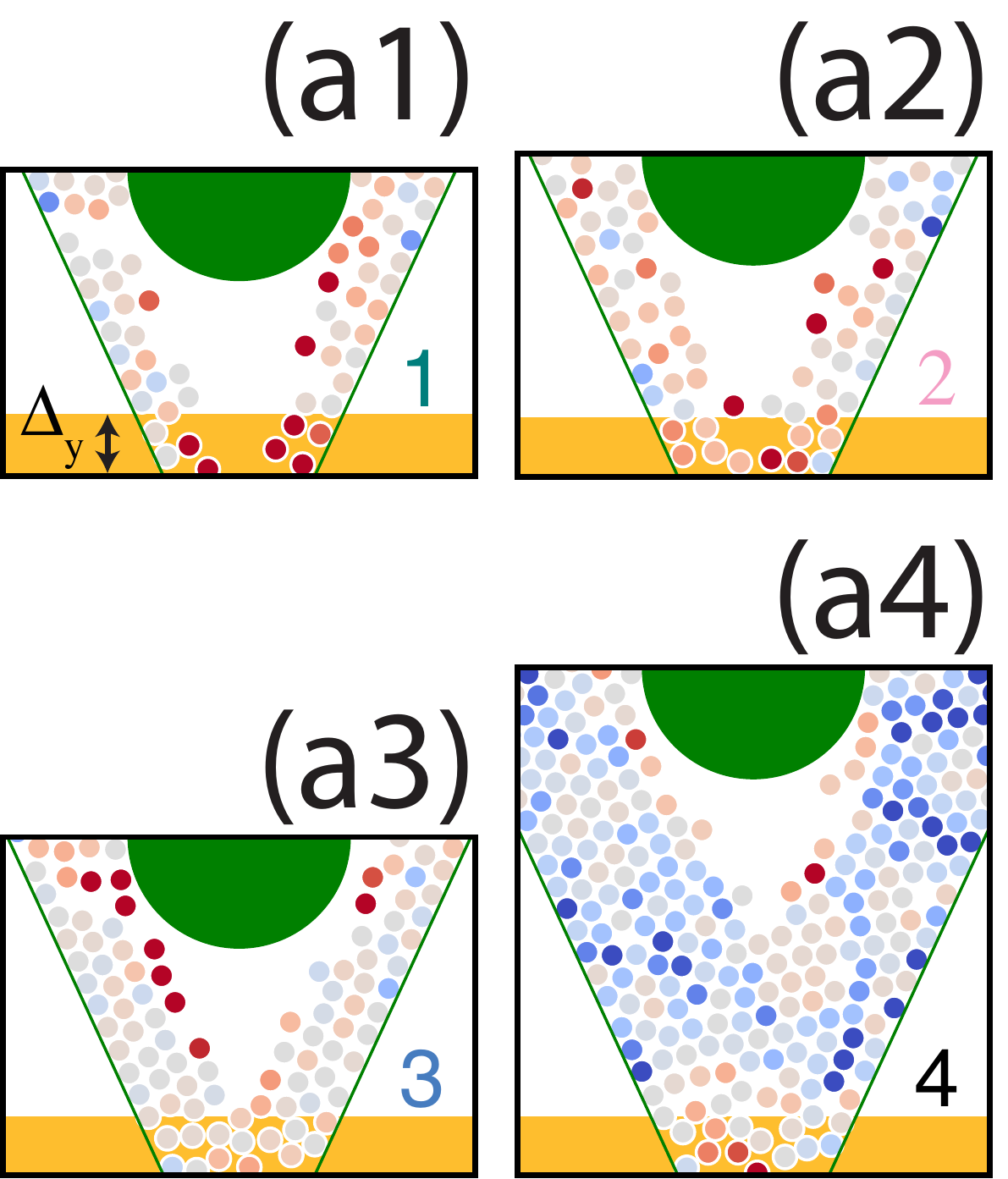}
\caption{\label{fig:factorized_flowrate1_snapshots} (Color online)
  (a1-a3) Representative snapshots of the system, focusing on the zone
  below the obstacle, placed at 1, 2, and 3 as labeled in
  Fig. \ref{fig:factorized_flowrate1}(a). (a4) Another snapshot where
  the obstacle is placed much further away from the exit of the hopper
  with $H/d \approx 21.414$. Particles are colored by the blue-red
  scheme in Fig. \ref{fig:tetris_model}(a), and their properties at
  the hopper exit are measured within the yellow region of height
  $\Delta_y=0.03L$.}
\end{figure}

Based on Fig. \ref{fig:factorized_flowrate1}, our explanation of the
flow rate enhancement by the concentration mechanism goes as
follows. Before the peak, $\phi_l^E<\phi_o^c$, and the position of the
obstacle controls the bottleneck flux $J_a^{BN}$ arriving at hopper
exit. The exit adjusts its $\phi_l^E$ in order to maintain the steady
state $J_a^E$, that is, to maintain the flux balance
$J_a^{BN}=J_a^E$. In this regime, what happens at the hopper exit does
not affect the flux $J_a^{BN}$. A representative snapshot is shown in
Fig. \ref{fig:factorized_flowrate1_snapshots}(a1), where we can see
that there are few particles at the exit because $\phi_l^E$ is low. On
the other hand, at or after the peak, $\phi_l^E \geq \phi_o^c$, as
shown in Fig. \ref{fig:factorized_flowrate1_snapshots}(a2) and
(a3). The particles with $\phi_l^E \geq \phi_o^c$ at the hopper exit
work together like a plug blocking the flow, and therefore
$J_a^{BN}>J_a^E$. In this situation, the steady state with no coupling
between the downstream exit and the upstream bottleneck regions is not
possible. If we further place the obstacle at a higher position, the
area packing fraction in the entire region between the obstacle and
the exit will build up, until the particles block the flow through the
bottleneck created by the obstacle, as shown in
Fig. \ref{fig:factorized_flowrate1_snapshots}(a4). The plateau of
$v_y^{BN}$ is an initial response to the downstream-upstream
(exit-bottleneck) coupling. Then $v_y^{BN}$ drops after the flow rate
peak, similar to a transition from the free-flow regime to the
congested-flow regime where the waiting room effect occurs
\cite{alonso-marroquin12}.

Finally, a hopper discharging relatively strongly-driven particles
with a larger $\alpha_y=0.439$ exhibits no flow rate enhancement. This
is because the converging geometry of the hopper cannot effectively
concentrate particles with a high $v_y^E/v_{yo}^E>1$ while discharging
them and $\phi_l^E$ never becomes higher than $\phi_o^c$ to make a
flow rate peak. This is inspected similarly in
Fig. \ref{fig:factorized_flowrate2}(a, b1) in the next section, where
we also propose an artificial merging strategy by utilizing the
concentration mechanism to create a man-made flow rate enhancement
peak.

\subsection{The importance of particle concentration mechanism within the space below the obstacle}
\label{Effect_merge_dependence}
\begin{figure}
\includegraphics[width=0.40\textwidth]{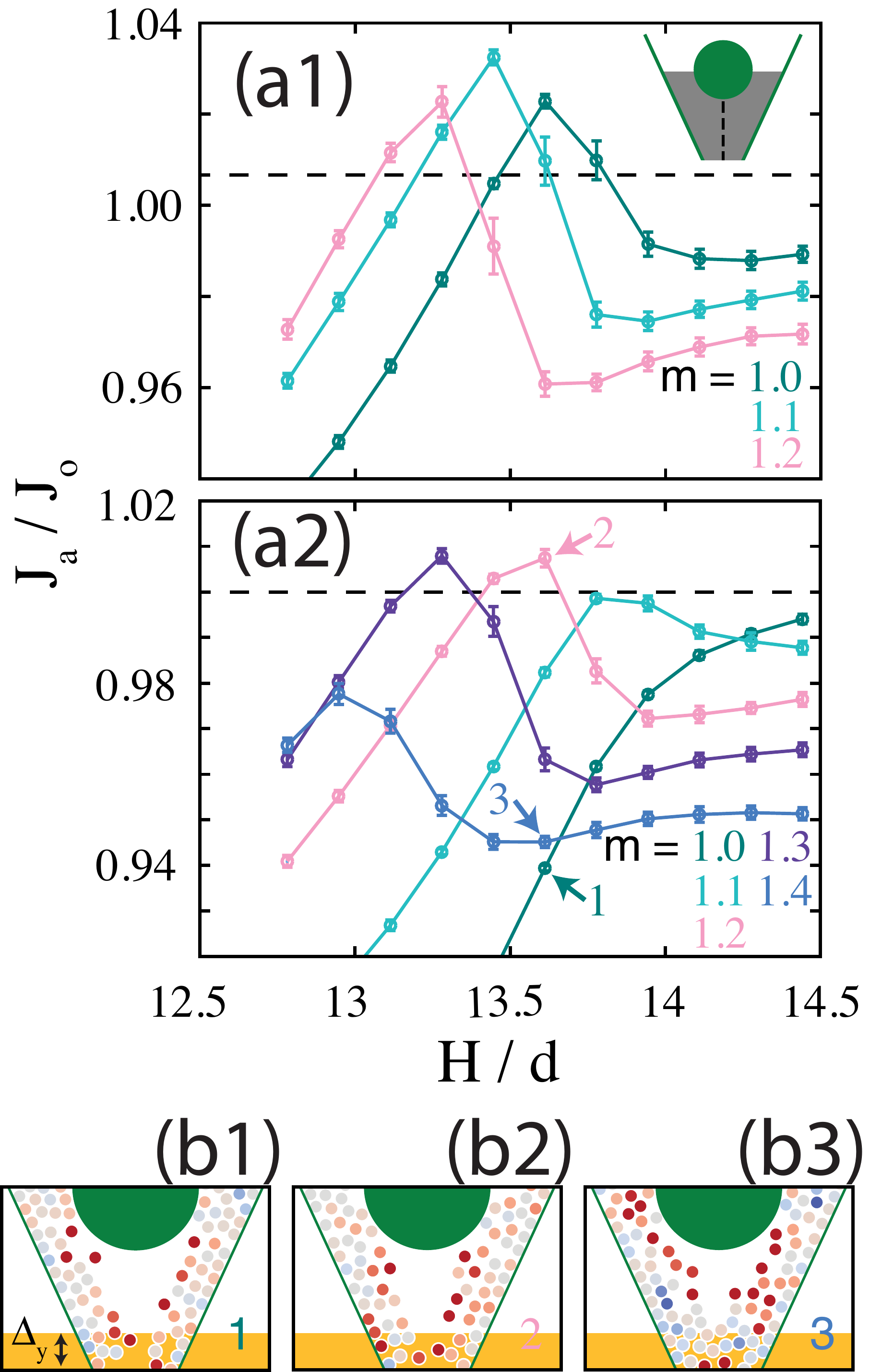}
\caption{\label{fig:merge_dependence1} (Color online) (a1-a2)
  Normalized hopper flow rates $J_a/J_o$ measured at the exit of a
  hopper with $\theta_2=\theta_1$, containing a round obstacle and
  $N=2048$. The driving strength $\alpha_{x}=m$ is $1.0$ (dark green),
  $1.1$ (cyan), $1.2$ (pink), $1.3$ (dark purple), and $1.4$ (navy) if
  particles in the shaded zone in-between $y=[0,H]$ move horizontally
  closer to the centerline (vertical dashed line) of the hopper, as
  shown by the inset. Otherwise, $\alpha_{x}=1.0$. The vertical
  driving strength $\alpha_y=0.333$ in (a1) and $0.439$ in (a2),
  respectively. The speed-up rate $r_s=1.0$. (b1-b3) Representative
  snapshots of the system, similar to
  Fig. \ref{fig:factorized_flowrate1_snapshots} but representing the
  three cases indicated by arrows in (a2).}
\end{figure}

To further validate the effect of particle concentration mechanism
within the domain below the obstacle on the flow rate, we artificially
compel particles to merge horizontally on their way toward the hopper
exit when they have moved to the shaded zone marked in the inset of
Fig. \ref{fig:factorized_flowrate1}(b). We achieve this by introducing
an anisotropic driving strength $\alpha_x = m \ge 1.0$ for particles
in the domain in-between $y=[0,H]$ that encourages them to attempt
moving horizontally towards the centerline of the hopper. On the other
hand, if particles in the same domain attempt to move horizontally
away from the centerline of the hopper, $\alpha_x = 1.0$ as usual. In
summary, ${\alpha _x} = \{ \begin{array}{*{20}{c}}
  {m,\,\,\,{\rm{towards}}\,{\rm{the}}\,{\rm{centerline}}}\\ {1,\,\,\,{\rm{otherwise}}\,\,\,\,\,\,\,\,\,\,\,\,\,\,\,\,\,\,\,\,\,\,\,\,\,\,\,\,\,}
\end{array}$. This artificial merging strategy presumably improves
concentrating efficiency beneath the obstacle by guiding particles to
where more space is available and reducing their probability of
hitting the hopper walls. We find that applying this strategy can
generate a higher flow rate peak of weakly-driven particles or can
even achieve flow rate enhancement in a relatively strongly-driven
system that originally exhibits no such phenomenon, as shown in
Fig. \ref{fig:merge_dependence1}(a1) and (a2), respectively.

\begin{figure}
\includegraphics[width=0.40\textwidth]{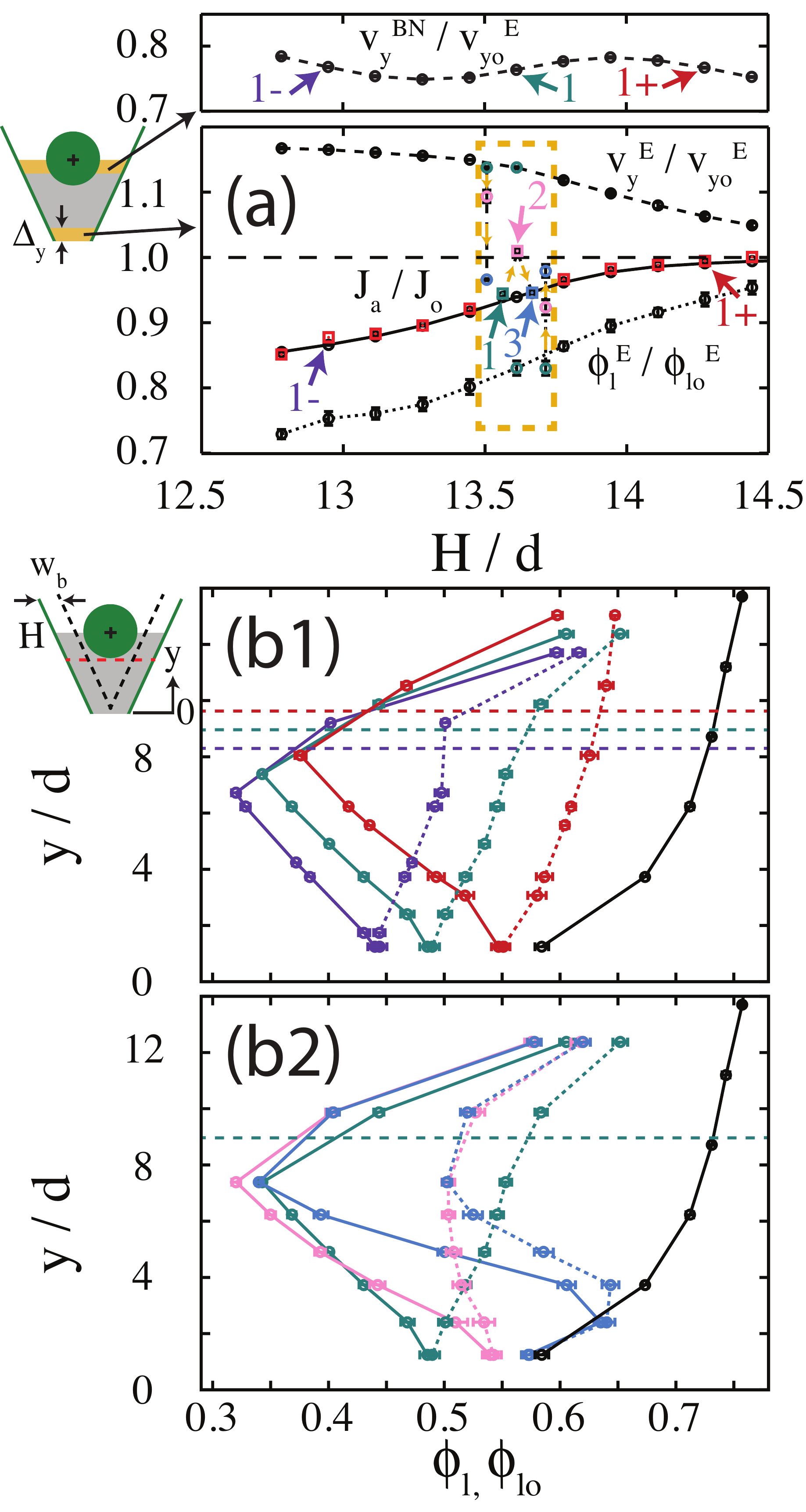}
\caption{\label{fig:factorized_flowrate2} (Color online) (a) The same
  plot as Fig. \ref{fig:factorized_flowrate1}(a) with $N=2048$ and
  $\alpha_y=0.439$ but different $\alpha_{x}=m$ to clarify the
  artificial merging strategy. Three selected cases with $m=1.0$ at
  different $H/d$ are labeled as 1- (dark purple), 1 (dark green), and
  1+ (red) while another two cases with $m=1.2$ and $1.4$ at $H/d$ of
  case 1 (= $13.612$) are labeled as 2 (pink) and 3 (navy),
  respectively. The data for cases 1, 2, and 3 in the dashed box are
  $v_y^E/v_{yo}^E$, $J_a^E/J_o^E$ on top of duplicate $J_a/J_o$, and
  $\phi_l^E/\phi_{lo}^E$ from left to right (shifted horizontally for
  better visibility). The orange arrows indicate the direction of
  increasing $m$. The corresponding normalized hopper flow rate
  $J_a/J_o$ with $m=1.0$ (solid line) and $m=1.2$ and $1.4$ (circles)
  are copied from Fig. \ref{fig:merge_dependence1}(a2) for
  reference. (b1) The same plot as
  Fig. \ref{fig:factorized_flowrate1}(b), except for cases 1-, 1, and
  1+. (b2) The same plot as (b1), except for cases 1, 2, and 3.}
\end{figure}

\begin{figure}
\includegraphics[width=0.40\textwidth]{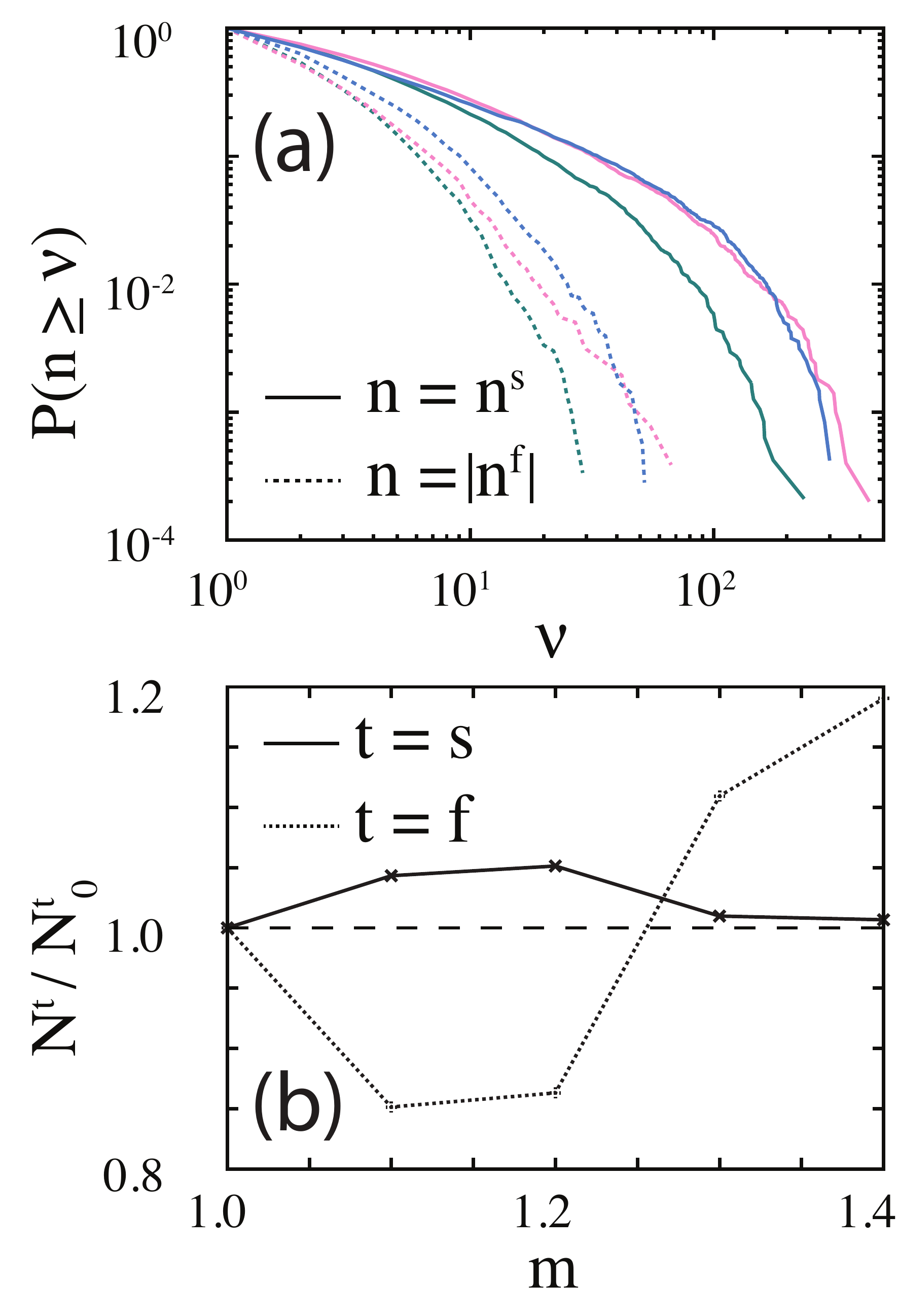}
\caption{\label{fig:merge_dependence2} (Color online) Representative
  (a) $P(n \ge \nu)$, using log-10 scales for both axes, and (b)
  $N^t/N^t_0$ in the shaded zone in-between $y=[0,H]$, as shown in the
  inset of Fig. \ref{fig:factorized_flowrate2}(b1), of the three
  selected cases, where a round obstacle is placed at $H/d = 13.612$,
  indicated by arrows in Fig. \ref{fig:merge_dependence1}(a2) with $m
  = 1.0$ (dark green), $1.2$ (pink), and $1.4$ (navy).}
\end{figure}

In Fig. \ref{fig:merge_dependence1}(a1), we plot the normalized flow
rate $J_a/J_o$ as a function of $H/d$ using the same $\alpha_y =
0.333$. When $m = 1.0$, the system exhibits the same flow rate peak
without the artificial merging strategy, as shown in
Fig. \ref{fig:N_dependence_J}(a1) and
Fig. \ref{fig:factorized_flowrate1}(a). We can maximize the peak value
by using an optimal $m = 1.1$. In
Fig. \ref{fig:merge_dependence1}(a2), we plot $J_a/J_o$ with an
amplified $\alpha_y = 0.439$ and the system exhibits no flow rate peak
when $m = 1.0$ within the same range of $H/d$. However, we can see
that $J_a/J_o$ exhibits a local peak shifting to the left as $m$ is
increased. The peak value of $J_a/J_o$ becomes larger than unity as
$m$ falls between $1.2$ and $1.3$. We argue that using $m = 1.2$
decreases the probability of particles below the obstacle hitting the
hopper walls, which allows them to merge more smoothly on their way
towards the hopper exit in the available space near the centerline,
promoting a higher maximum flow rate as they accumulate. Further
increasing the value of $m$ to $1.4$ lowers the enhanced flow rate
because particles begin to experience more failed moves near the
centerline of the hopper due to more frequent interparticle
interference and can even clog at its exit, yielding less effective
concentration mechanism. We first verify the above argument visually
using the representative snapshots of the selected cases shown in
Fig. \ref{fig:merge_dependence1}(b1) - (b3), where particles at the
hopper exit experience the characteristic transition from $\phi_l^E <
\phi_o^c$ to $\phi_l^E > \phi_o^c$, a similar scenario as in
Fig. \ref{fig:factorized_flowrate1_snapshots}. Next, we put our
arguments under quantitative investigation and show the results in
Fig. \ref{fig:factorized_flowrate2} and
Fig. \ref{fig:merge_dependence2}.

The artificial merging strategy can be interpreted as a way of
utilizing the concentration mechanism in Sec. \ref{peaking_mechanism}
to create a flow rate enhancement peak. To show this, we plot
$J_a/J_o$ with $\alpha_y=0.439$ and $m = 1.0$, copied from
Fig. \ref{fig:merge_dependence1}(a2), together with its measured
decomposed variables ${{\phi _l^E} \mathord{\left/ {\vphantom {{\phi
          _l^E} {\phi _{lo}^E}}} \right.  \kern-\nulldelimiterspace}
  {\phi _{lo}^E}}$ and ${{v_y^E} \mathord{\left/ {\vphantom {{v_y^E}
        {v_{yo}^E}}} \right.  \kern-\nulldelimiterspace} {v_{yo}^E}}$,
and the calculated $J_a^E/J_o^E$ in the lower box of
Fig. \ref{fig:factorized_flowrate2}(a). As in
Fig. \ref{fig:factorized_flowrate1}(a), $\phi _l^E/\phi _{lo}^E$ and
$v_y^E/v_{yo}^E$ at the hopper exit monotonically increases and
decreases, respectively, with increasing $H/d$ as the obstacle is
placed further away from the hopper exit. However, $\phi _l^E$ is
never greater than $\phi_o^c$ while maintaining a high $v_y^E$ to
create the characteristic transition shown in
Fig. \ref{fig:factorized_flowrate1}(a), and therefore we do not
observe an enhanced flow rate. Now, we switch on the artificial
merging strategy by increasing $m=1.0$ to $m=1.2$ and $1.4$, labeled
as cases 1, 2, and 3 in the dashed box in
Fig. \ref{fig:factorized_flowrate2}(a) (corresponding to identically
labeled cases in Fig. \ref{fig:merge_dependence1}(a2)). Together, we
mark how $J_a^E/J_o^E$ and the associated ${{\phi _l^E}
  \mathord{\left/ {\vphantom {{\phi _l^E} {\phi _{lo}^E}}} \right.
    \kern-\nulldelimiterspace} {\phi _{lo}^E}}$ and ${{v_y^E}
  \mathord{\left/ {\vphantom {{v_y^E} {v_{yo}^E}}} \right.
    \kern-\nulldelimiterspace} {v_{yo}^E}}$ change with $m$, indicated
by the yellow arrows in the figure. Using $m = 1.2$ successfully
increases ${{\phi _l^E} \mathord{\left/ {\vphantom {{\phi _l^E} {\phi
          _{lo}^E}}} \right.  \kern-\nulldelimiterspace} {\phi
    _{lo}^E}}$ while ${{v_y^E} \mathord{\left/ {\vphantom {{v_y^E}
        {v_{yo}^E}}} \right.  \kern-\nulldelimiterspace} {v_{yo}^E}}$
is only mildly reduced at the hopper exit. As a result, we observe
$J_a^E/J_o^E>1$. A larger $m = 1.4$ further increases ${{\phi _l^E}
  \mathord{\left/ {\vphantom {{\phi _l^E} {\phi _{lo}^E}}} \right.
    \kern-\nulldelimiterspace} {\phi _{lo}^E}}$ but simultaneously
${{v_y^E} \mathord{\left/ {\vphantom {{v_y^E} {v_{yo}^E}}} \right.
    \kern-\nulldelimiterspace} {v_{yo}^E}}$ becomes too small to make
$J_a^E/J_o^E>1$. Again, the calculated $J_a^E/J_o^E$ agrees with the
measured $J_a/J_o$ well in all studied cases. In the upper box of
Fig. \ref{fig:factorized_flowrate2}(a), we plot $v_y^{BN}/v_{yo}^E$,
which exhibits a gentle local peak. The peak indicates a weak
downstream-upstream coupling. Nevertheless, the overall effect of the
concentration mechanism on relatively strongly-driven particles is not
significant enough to create a flow rate enhancement peak.

To learn more about the variation of $\phi_l$ under the obstacle
placed at different $H/d$ without the artificial merging strategy
($m=1.0$), we select three $H/d$, labeled as 1-, 1, and 1+ in
Fig. \ref{fig:factorized_flowrate2}(a), and plot their $\phi_l$ in
Fig. \ref{fig:factorized_flowrate2}(b1). In all the cases, $\phi_l$
first decreases until somewhere below the obstacle and then increases
until the hopper exit. Plotting $\phi _l$ next to the hopper wall
reveals that the densest part of the local area packing fraction in
this region decreases monotonically towards the hopper exit, which
indicates that the converging geometry of the hopper does not
effectively help to collect particles on their way out. Its value is
even smaller than when the hopper contains no obstacle. While the
particles move fast towards the hopper exit because of a large
$\alpha_y$, particles with a decaying $\phi_l$ passing the obstacle
travel even faster so long as the concentrating capability of the
hopper fails to keep $\phi _l$ next to the hopper wall constant, as is
the case in Fig. \ref{fig:factorized_flowrate1}(b), which reciprocally
further reduces $\phi _l$ next to the hopper wall below the
obstacle. Therefore, ${{\phi _l^E} \mathord{\left/ {\vphantom {{\phi
          _l^E} {\phi _{lo}^E}}} \right.  \kern-\nulldelimiterspace}
  {\phi _{lo}^E}}$ cannot reach a high enough value with a still fast
${{v_y^E} \mathord{\left/ {\vphantom {{v_y^E} {v_{yo}^E}}} \right.
    \kern-\nulldelimiterspace} {v_{yo}^E}}$ to create a flow rate
enhancement.

In Fig. \ref{fig:factorized_flowrate2}(b2), we plot $\phi_l$ under the
obstacle with a fixed $H/d=13.612$ but different strengths of the
artificial merging strategy with $m = 1.0$, $1.2$, and $1.4$, labeled
as 1, 2, and 3 in Fig. \ref{fig:merge_dependence1}(a2) or
Fig. \ref{fig:factorized_flowrate2}(a). The value of $\phi_l$ again
decreases monotonically from the narrowest points between the obstacle
and the two hopper walls to somewhere below the obstacle. Compared
with particles subject to an isotropic horizontal $\alpha_x=m=1$,
$\phi_l$ of particles with $m = 1.2$ is smaller in the vicinity below
the obstacle but soon grows above all the way towards the exit. The
rise of $\phi_l$ is more effective than the slight drop of $v_y$ to
lead to the flow rate enhancement. When we overdo the artificial
merging strategy with an $m=1.4$, $\phi_l$ grows much more severely
towards the exit and even exhibits a sudden drop before the exit,
destroying the flow rate enhancement, similar to the after-peak case 3
presented in Fig. \ref{fig:factorized_flowrate1}(b). In addition, we
plot $\phi_l$ next to the hopper wall with the dotted lines, whose
value with $m = 1.0$ decreases monotonically, as we have seen
previously. In contrast, both the cases with $m = 1.2$ and $1.4$
decrease only initially but then increase, but a sudden drop close to
the exit is detected when overdoing the artificial merging strategy
with $m=1.4$. We notice that the nonlinear variation of $\phi_l$ near
the hopper exit with increasing $m$ closely resembles the trend found
before, at, and after an enhanced flow rate peak by moving the
obstacle up in a hopper discharging isotropic particles, as shown in
Fig. \ref{fig:factorized_flowrate1}(b).

Focusing on the particle data with an anisotropic $\alpha_x$ collected
in the shaded space in the inset of
Fig. \ref{fig:factorized_flowrate2}(b1), we go one step further to
quantitatively examine the complementary cumulative distribution
function $P(n \ge \nu)$ with respect to $m$, which gives the
probability of randomly finding an $n$ no smaller than $\nu$
\cite{newman09}. The value of $n$ represents particles that
successfully or unsuccessfully update their positions $n^s$ or $\left|
{{n^f}} \right|$ times. We perform the calculation using $990,000$
position-update cycles after a system forgets its initial state. We
also calculate the corresponding $N^t/N^t_0$, where $N^t$ is the total
number of successful-type particles $N^s$ or failed-type particles
$N^f$ counted while building $P(n^s \ge \nu)$ or $P(\left| {{n^f}}
\right| \ge \nu)$, and $N^t_0$ is the same quantity with $m = 1.0$
used for normalization. The resulting $P(n \ge \nu)$ of the three
selected cases with the same $H/d$, labeled as 1, 2, and 3 in
Fig. \ref{fig:merge_dependence1}(a2) or
Fig. \ref{fig:factorized_flowrate2}(a), are shown in
Fig. \ref{fig:merge_dependence2}(a). As $m$ increases from $1.0$ to
$1.2$, the range of $\nu$ with $P(\left| {{n^f}} \right| \ge \nu)$
between $0.1$ and $1.0$ barely expands, meaning that about $90\%$ of
the motionless particles do not experience a higher failure rate of
updating their positions. On the other hand, $P(n^s \ge \nu)$ shows an
overall increase which leads to the improved flow rate. As $m$
increases again from $1.2$ to $1.4$, $P(n^s \ge \nu)$ shows no
significant increase while $P(\left|{{n^f}} \right| \ge \nu)$ notably
increases, resulting in the drop of the flow rate. The plot of
$N^t/N^t_0$ against $m$ directly confirms this observation, as shown
in Fig. \ref{fig:merge_dependence2}(b). Compared with the case of $m =
1.0$, the case of $m = 1.2$ has its number of failed particles $N^f$
reduced by about $15\%$ and number of successful particles $N^s$
increased by about $5\%$. These numbers are also superior to those of
the case with $m = 1.4$, which have almost no improvement on $N^s$ but
a dramatic increase in $N^f$ by almost $20\%$.

\subsection{The sensitivity of particle concentration mechanism within the space below the obstacle}
\label{Effect_merge_sensitivity}
\begin{figure}
\includegraphics[width=0.40\textwidth]{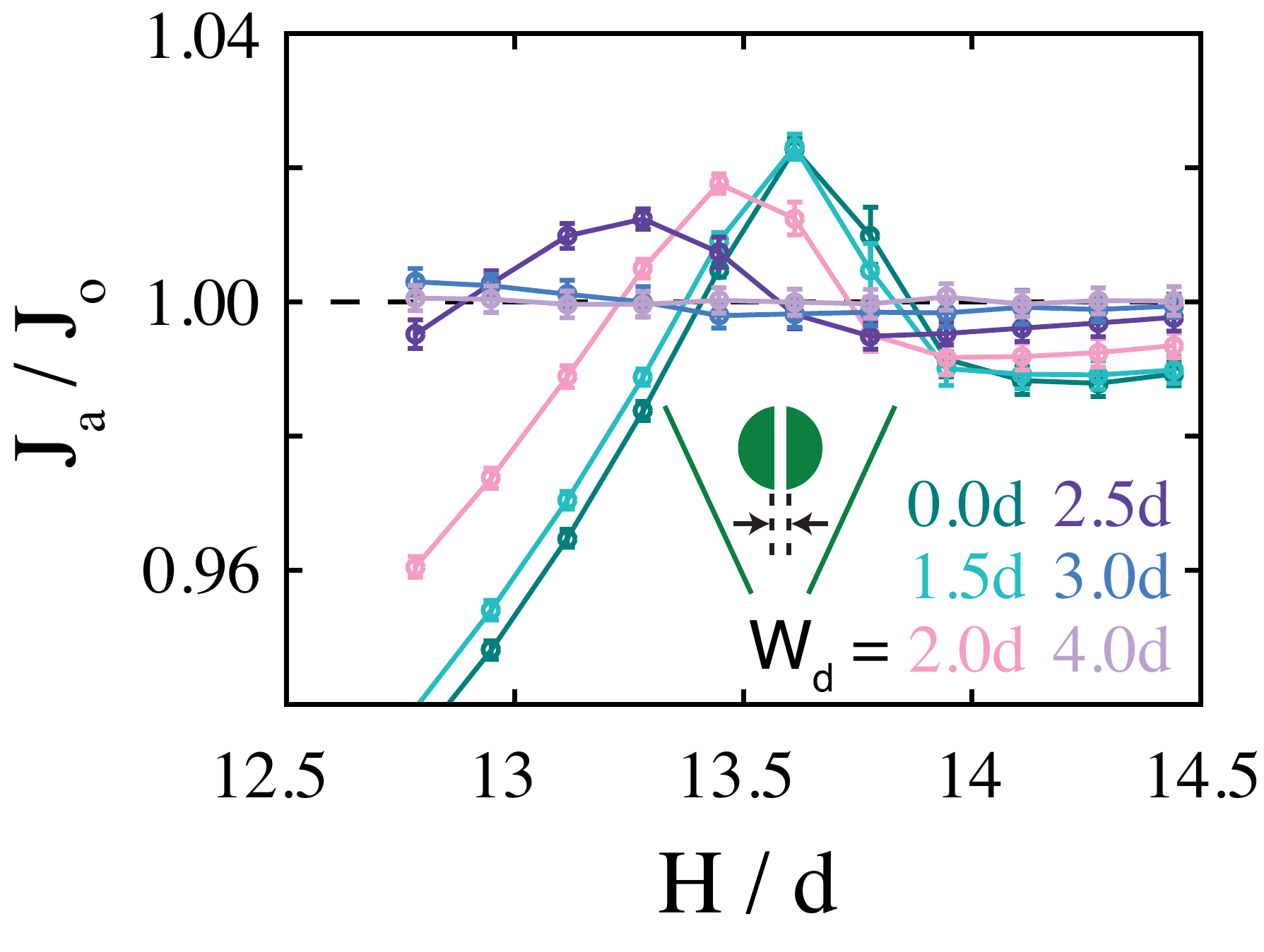}
\caption{\label{fig:duct_dependence} (Color online) Normalized hopper
  flow rates $J_a/J_o$ measured at the exit of a hopper with
  $\theta_2=\theta_1$, and containing a round obstacle with a hollow
  duct of width $W_d=0.0d$ (dark green), $1.5d$ (cyan), $2.0d$ (pink),
  $2.5d$ (dark purple), $3.0d$ (navy), and $4.0d$ (purple). The total
  number of particles in the system $N=2048$, the driving strengths
  $\alpha_{x}=1.0$, $\alpha_y=0.333$, and the speed-up rate
  $r_s=1.0$.}
\end{figure}

To understand if simply filling the available space under the obstacle
can also improve the flow rate, we introduce a different obstacle with
a hollow duct of width $W_d>d$. The hollow duct offers a shortcut that
allows particles to arrive at the available space below the obstacle
faster, potentially leading to a more efficient particle concentration
mechanism. The results are shown in
Fig. \ref{fig:duct_dependence}. Counterintuitively, this approach has
very limited effect on improving the flow rate, except when the duct
width is very narrow with $W_d=1.5$. Much wider duct widths
($W_d>1.5$) that allow more particles through can even have a negative
influence on the enhanced flow rate. This approach probably hinders
the original particle concentrating process due to particle exclusion
upon collision. Additionally, particles passing through the uniform
duct are subject to little concentrating assistance. Assigning an
anisotropic $\alpha_x$ to particles below the obstacle, on the other
hand, is a more effective strategy which improves the concentrating
efficiency by directly reducing the collision probability between
particles and the hopper.

\subsection{Investigating factors related to the waiting room effect}
\label{Effect_waiting_room}
The waiting room effect, wherein particles are first slowed down by
the obstacle and then speed up due to the external driving force
within the triangular-ish void space between the obstacle and the
hopper exit, has been suggested to be responsible for the enhanced
flow rate. Here we examine the factors related to the waiting room
effect with the Tetris-like model by looking at its components one by
one, namely the size of the void space, particle speed-up rate, and
exit geometry of the hopper.

\subsubsection{The size of the waiting room}
\label{Effect1_waiting_room}
\begin{figure}
\includegraphics[width=0.40\textwidth]{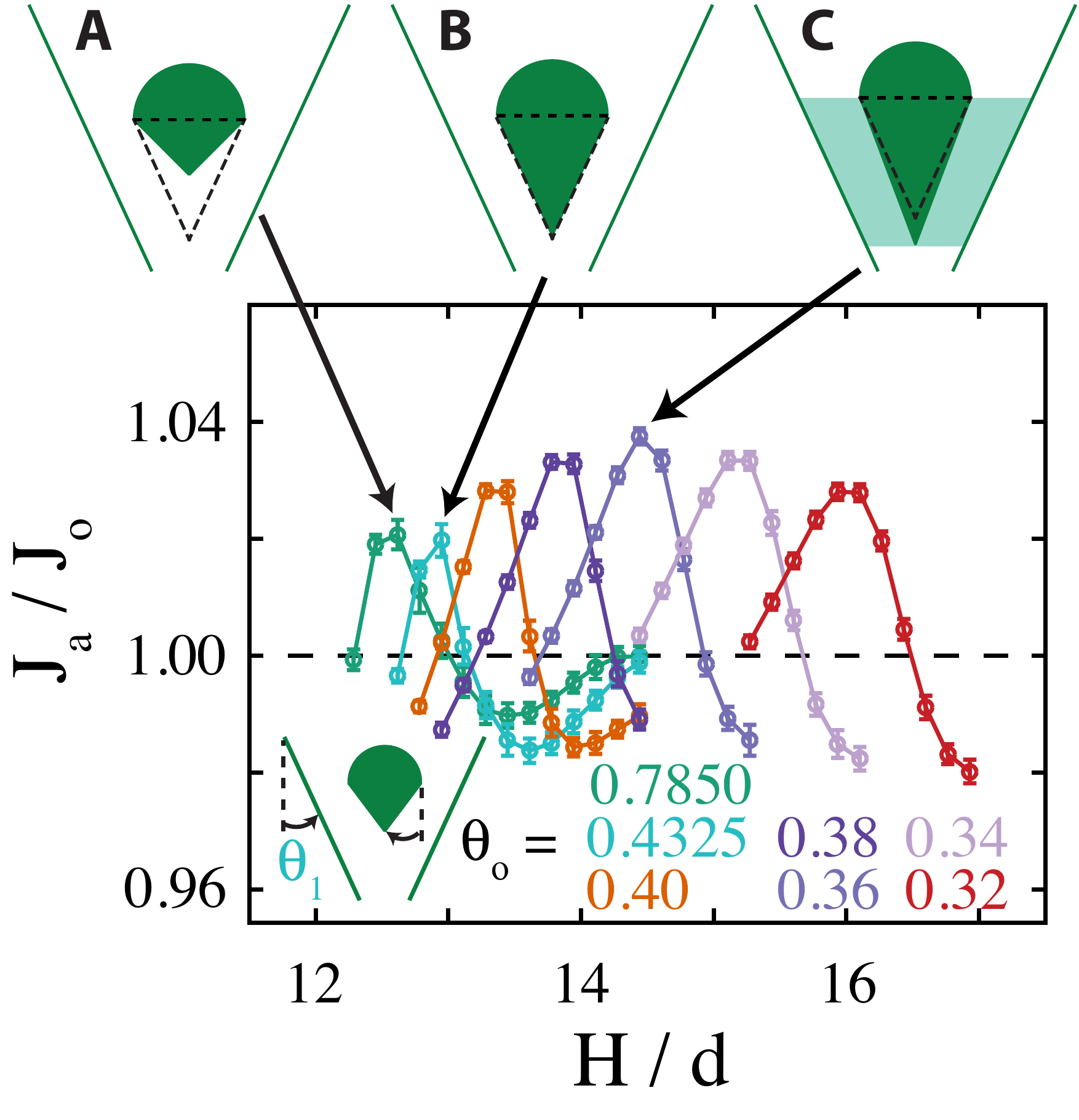}
\caption{\label{fig:awl_dependence_1} (Color online) Normalized hopper
  flow rates $J_a/J_o$ measured at the exit of a hopper with
  $\theta_2=\theta_1=0.4325$, and containing an obstacle composed of a
  semicircle and an isosceles triangle with an area $A_t$ and an angle
  to the vertical $\theta_o=0.7850$ (dark green), $0.4325$ (cyan),
  $0.40$ (pumpkin), $0.38$ (dark purple), $0.36$ (navy), $0.34$
  (purple), and $0.32$ (red). The total number of particles in the
  system $N=2048$, the driving strengths $\alpha_{x}=1.0$,
  $\alpha_y=0.333$, and the speed-up rate $r_s=1.0$. The insets are
  three representative system setups whose $J_a/J_o$ is at a peak
  value (denoted by the arrows), with the net waiting room area
  $A_w=A_w^m-A_t>0$ (\textbf{A}), $=0$ (\textbf{B}), and $<0$
  (\textbf{C}), where $A_w^m$ is the maximum waiting room area (dashed
  triangles). The light green area in (\textbf{C}) indicates the
  shrinking geometry from the obstacle to the hopper exit.}
\end{figure}

To understand the effect of the triangular-ish void space on the
enhanced flow rate, we replace the round obstacle by one composed of
an identical upper semicircle and a lower triangle with an area $A_t$
and an angle $\theta_o$ measured from the vertical, as shown in
Fig. \ref{fig:awl_dependence_1}. We keep the shape of the upper half
of the obstacle unchanged so that we can focus on the size of the
waiting room, which is controlled by the lower half of the
obstacle. When $\theta_o=\pi/2$, $A_t=0$. We define the maximum
waiting room area $A_w^m$ as the space circumscribed by the lower
boundary of the semicircle and the two lines parallel to the hopper
walls, as indicated by the dashed triangle in the insets of
Fig. \ref{fig:awl_dependence_1}. The net waiting room area is found by
deducting the area occupied by the lower triangle of the obstacle,
$A_w=A_w^m-A_t$. It decreases with decreasing $\theta_o$ because of
the associated increase of $A_t$. We tested seven different values of
$\theta_o$ between $0.7850$ and $0.32$. When $\theta_o=0.7850$, the
area of the composite obstacle can be fully encompassed by the round
obstacle used before. We monitored the corresponding normalized flow
rate $J_a/J_o$ and observed that a flow rate peak appears in all the
tested cases. We can see that the value of the flow rate peak
decreases slightly as $A_w$ changes from positive with
$\theta_o=0.7850$ to zero with $\theta_o=\theta_1=0.4325$, showing
that the waiting room size affects the enhanced flow rate but is not
essential to it. Surprisingly, as we further decrease $\theta_o$ to
about $0.36$ the value of the flow rate peak increases and reaches an
optimum with $A_w<0$, where the widths of the two channels between the
obstacle and the hopper walls gradually shrink and particles must
become more concentrated while flowing through them. Similar results
with $A_w<0$ and $A_w>0$ have been reported experimentally and
numerically using frictional particles in a quasi-2D hopper governed
by Newtonian dynamics \cite{alonso-marroquin16}; however, here we test
the case of $A_w=0$ and leave all other factors untouched. It is clear
that eliminating the waiting room by reducing its area to zero does
not completely annihilate the enhanced flow rate peak but only affects
its magnitude.

\begin{figure}
\includegraphics[width=0.40\textwidth]{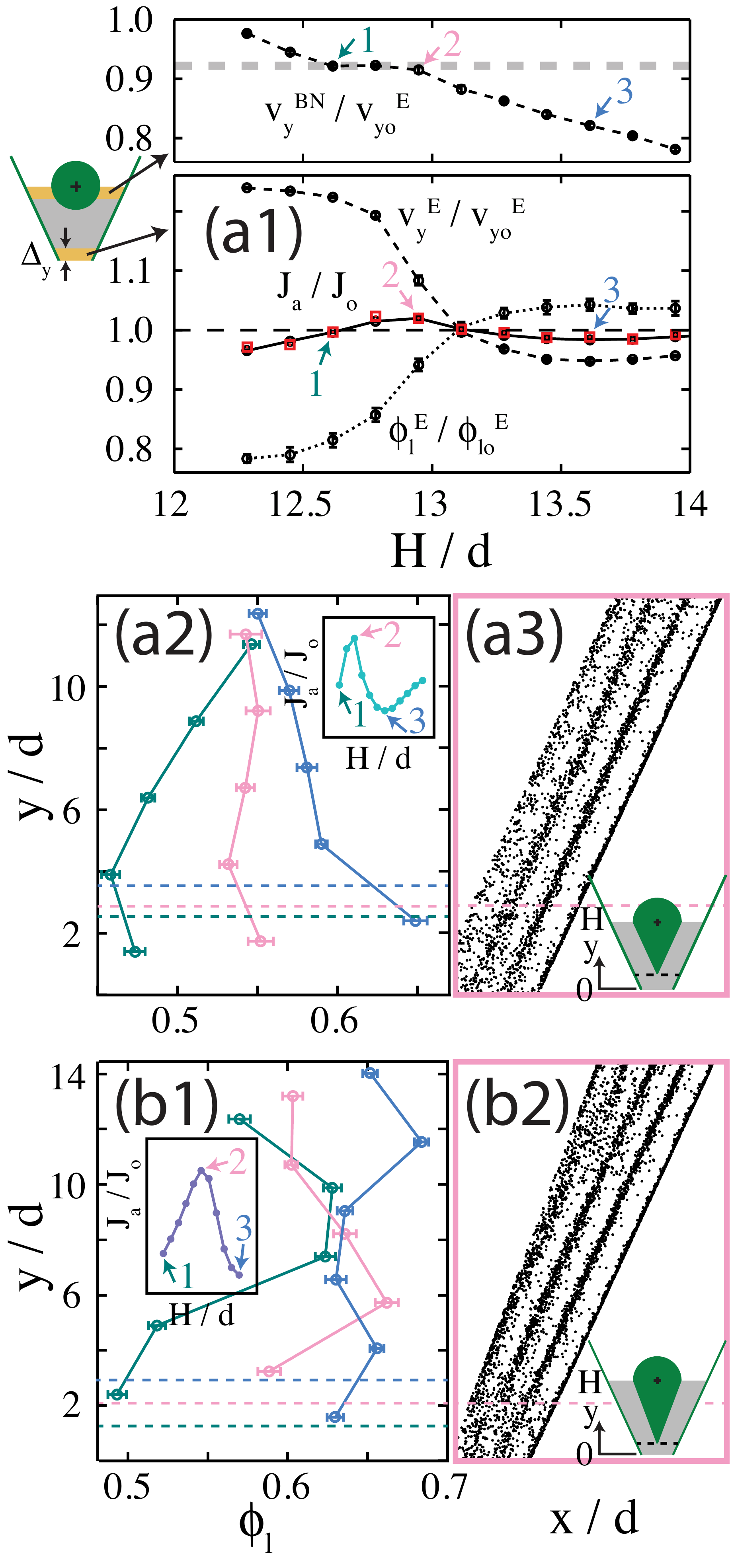}
\caption{\label{fig:awl_dependence_2} (Color online) a1) The same plot
  as Fig. \ref{fig:factorized_flowrate1}(a), except the waiting room
  area $A_w=0$ and $\theta_o=0.4325$. (a2) $\phi_l$ in-between
  $y=[0,H]$ for cases labeled as 1 (dark green), 2 (pink), and 3
  (navy) in (a1) and in the inset. The horizontal dashed lines
  indicate the lowest y positions of the obstacle. (a3) The
  corresponding instantaneous $x$ and $y$ position scatter plot of
  particles in the shaded zone with the same $y$ range for case 2 in
  (a2). The plot is obtained by overlapping $99$ snapshots, each
  separated by $10,000$ position-update cycles. (b1-b2) The same plots
  as (a2-a3), except $\theta_o=0.36$.}
\end{figure}

Our test shows that we can create an enhanced flow rate even though
the size of the waiting room is reduced to zero. To examine if the
mechanism in Sec. \ref{peaking_mechanism} can still explain the
phenomenon in this setup, we consider the lower box of
Fig. \ref{fig:awl_dependence_2}(a1). Here, we plot the measured
$J_a/J_o$ again with the waiting room area $A_w=0$ and
$\theta_o=0.4325$ (copied from Fig. \ref{fig:awl_dependence_1}), its
measured ${{\phi _l^E} \mathord{\left/ {\vphantom {{\phi _l^E} {\phi
          _{lo}^E}}} \right.  \kern-\nulldelimiterspace} {\phi
    _{lo}^E}}$ and ${{v_y^E} \mathord{\left/ {\vphantom {{v_y^E}
        {v_{yo}^E}}} \right.  \kern-\nulldelimiterspace} {v_{yo}^E}}$,
and the calculated $J_a^E/J_o^E$.

Our results show that $\phi _l^E/\phi _{lo}^E$ and $v_y^E/v_{yo}^E$ at
the hopper exit follow identical patterns of monotonically increasing
and decreasing, respectively, when the obstacle is placed further away
from the hopper exit, as we have seen in
Fig. \ref{fig:factorized_flowrate1}(a). Similarly, the flow rate peaks
where the characteristic transition in
Fig. \ref{fig:factorized_flowrate1}(a) happens. The results confirm
the validity of the particle concentration mechanism. In the upper box
of Fig. \ref{fig:awl_dependence_2}(a1), we plot $v_y^{BN}/v_{yo}^E$,
which exhibits a similar plateau to the one in
Fig. \ref{fig:factorized_flowrate1}(a). We also plot the corresponding
local area packing fraction $\phi_l$ under the obstacle in-between
$y=[0,H]$ in Fig. \ref{fig:awl_dependence_2}(a2). The variation of
$\phi_l$ becomes noticeably smaller but still visible after we
eliminate the waiting room area in
Fig. \ref{fig:factorized_flowrate1}(b). The visible variation of
$\phi_l$ may suggest interparticle spatial fluctuations, even though
the widths of the channels between the obstacle and the hopper walls
stay constant. Similar spatial fluctuations in packing fraction have
been reported both experimentally and numerically in a hopper or silo
discharging frictional granular materials
\cite{menon09,tewari13,durian16,zuriguel19}. Before or after the flow
rate peak, the value of $\phi_l$ decreases or increases monotonically
from the narrowest points between the obstacle and the two hopper
walls to somewhere above the lowest point of the obstacle. At the flow
rate peak, the variation of $\phi_l$ is the smallest but also the most
complicated. After the lowest point of the obstacle and immediately
before the hopper exit, $\phi_l$ increases in all three cases due to
the merging of the two channels and decreasing of the available space
to the particles. In addition, by plotting the instantaneous particle
$x$ and $y$ positions under the obstacle for the case of the maximum
flow rate enhancement, we can see the related fluctuation of $\phi_l$
on a particle scale, as shown in Fig. \ref{fig:awl_dependence_2}(a3).

Likewise, in Fig. \ref{fig:awl_dependence_2}(b1), we plot $\phi_l$
with $A_w<0$ and $\theta_o=0.36$. The shrinking geometry from the
obstacle to the hopper exit increases the degree of fluctuation of
$\phi_l$ which also reflects on the particle $x$ and $y$ position
scatter plot in Fig. \ref{fig:awl_dependence_2}(b2). These results of
$\phi_l$ with $A_w=0$ and $A_w<0$ show that as long as $\phi_l$ can
reach a value between $0.5$ and $0.6$ upon reaching the hopper exit, a
flow rate enhancement can be achieved and the intermediate fluctuation
of $\phi_l$ is not essential to the peaking phenomenon.

\subsubsection{The acceleration of particles within the waiting room}
\label{Effect2_waiting_room}
\begin{figure}
\includegraphics[width=0.40\textwidth]{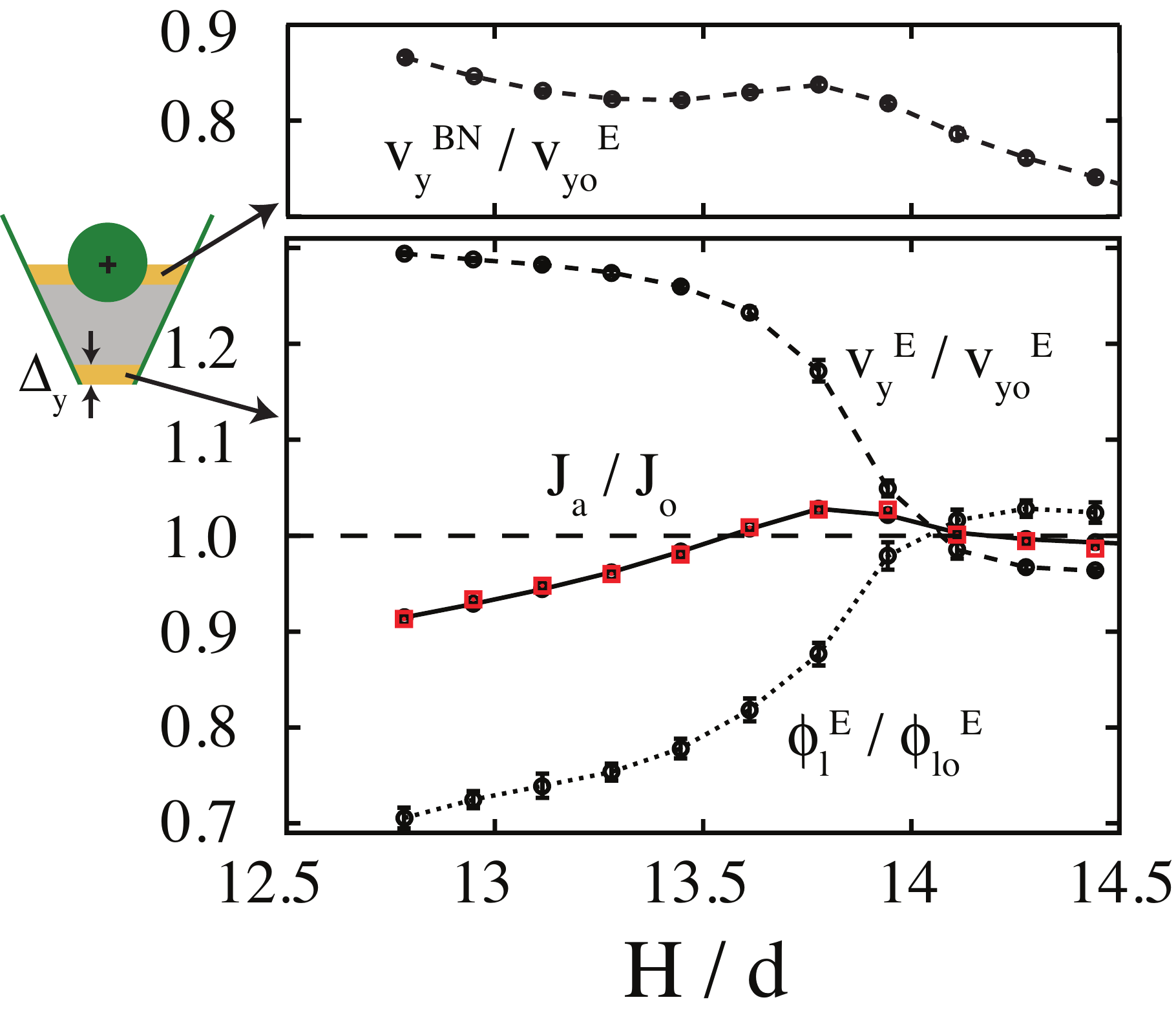}
\caption{\label{fig:factorized_flowrate1_rs_101} (Color online) The
  same plot as Fig. \ref{fig:factorized_flowrate1}(a) with
  $\alpha_{x}=1.0$, $\alpha_y=0.333$, and $N=2048$, but a larger
  $r_{s}=1.01$, which allows a particle successively updating its
  position to move farther during the next position-update cycle.}
\end{figure}

The second factor associated with the waiting room effect is the
acceleration of particles in the direction aligned with the external
driving force such as gravity. This factor is expressed by the term
$r_s^{n_i^s}$ in Eqn. \ref{tetris_algorithm}, where a greater than
unity $r_s$ is the speed-up rate that allows particle $i$, after
successfully updating its position $n_i^s$ times, to move farther
during the next position-update cycle. We test the concentration
mechanism in Sec. \ref{peaking_mechanism} with $r_s = 1.01$ and show
that it can still explain the flow rate enhancement even with particle
acceleration, as seen in
Fig. \ref{fig:factorized_flowrate1_rs_101}. Both $v_y^{BN}/v_{yo}^E$
and $J_a/J_o$ locally peak at the same value of $H/d$, similar to the
transition from the free-flow regime to the congested-flow regime
where the waiting room effect happens \cite{alonso-marroquin12}.

\begin{figure}
\includegraphics[width=0.40\textwidth]{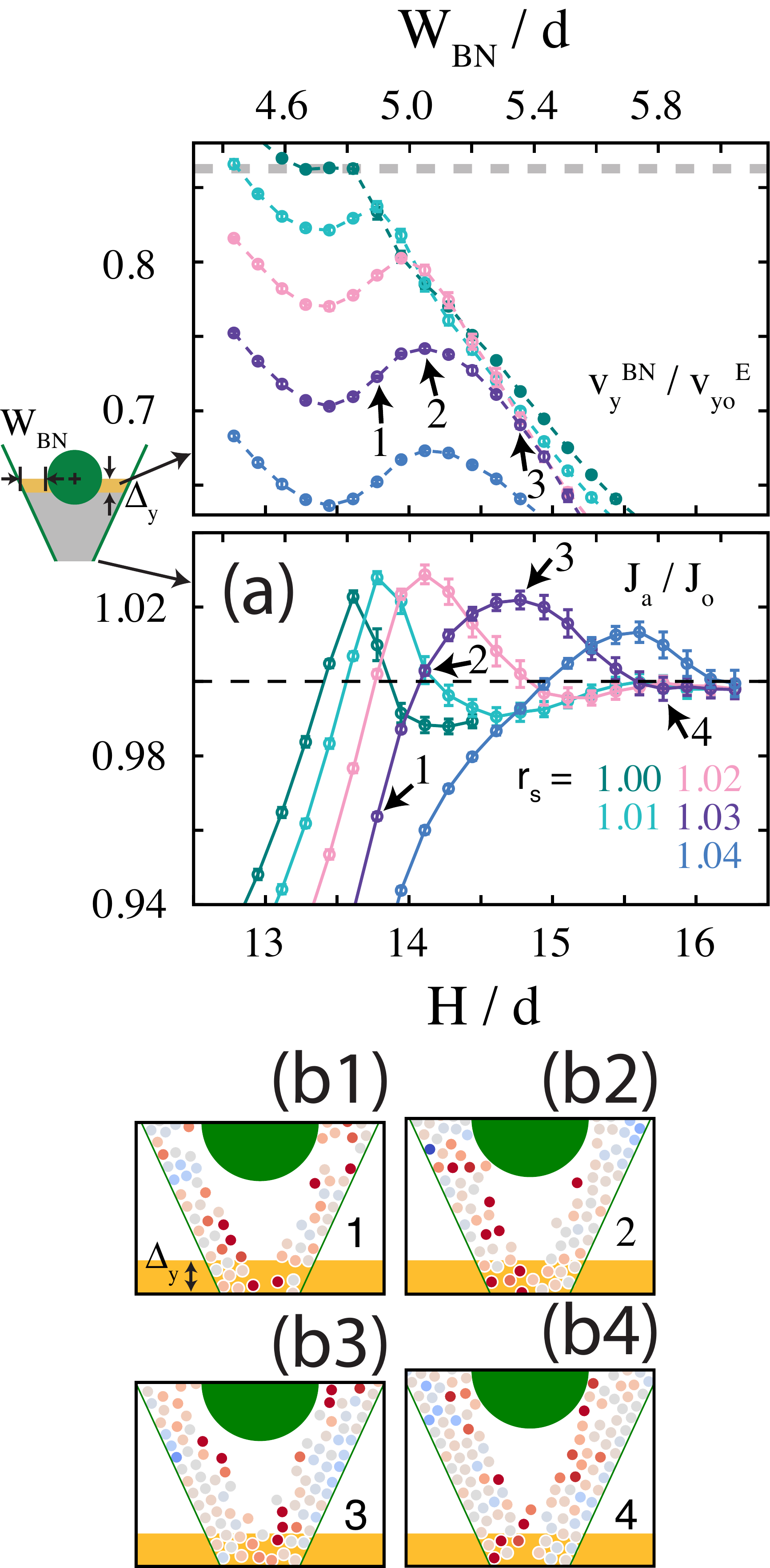}
\caption{\label{fig:speedup_dependence} (Color online) (a) The lower
  box shows a plot of normalized hopper flow rates $J_a/J_o$ measured
  at the exit of a hopper with $\theta_2=\theta_1$, containing a round
  obstacle. The upper box shows a plot of normalized averaged particle
  velocity $v_y^{BN}/v_{yo}^E$, measured at bottleneck $y_c = H -
  0.015L$ (yellow region in the inset). The driving strengths
  $\alpha_{x}=1.0$ and $\alpha_y=0.333$, and $N=2048$. The speed-up
  rate $r_s=1.00$ (dark green), $1.01$ (cyan), $1.02$ (pink), $1.03$
  (dark purple), and $1.04$ (navy). The lower and upper horizontal
  axes are normalized obstacle position $H/d$ and normalized
  bottleneck width $W_{BN}/d$, respectively. (b1-b4) Representative
  snapshots of the system, similar to
  Fig. \ref{fig:factorized_flowrate1_snapshots}, illustrating the four
  cases indicated by arrows in (a).}
\end{figure}

Including $r_s=1.01$, we also tested three other values of $r_s$ from
$1.02$ to $1.04$ and compared the obtained normalized flow rate
$J_a/J_o$ with that of $r_s=1.00$, the condition with no particle
acceleration. The results are shown in the lower box of
Fig. \ref{fig:speedup_dependence}(a). We can see that both the peak
value and the range of $H/d$ of the enhanced flow rate increase with a
milder rise of $r_s=1.01$ and $1.02$. However, using an even higher
particle speed-up rates towards to hopper exit ($r_s>1.02$) has a
negative impact on the flow rate peak value, as shown by the decreases
in the two measured quantities. This decline may also be attributed to
a higher collision frequency when more particles accelerate towards
the hopper exit. In the upper box of
Fig. \ref{fig:speedup_dependence}(a), we plot $v_y^{BN}/v_{yo}^E$,
which plateaus with $r_s=1.00$ or peaks with $r_s>1$. Four
representative snapshots with $r_s=1.03$ are shown in
Fig. \ref{fig:speedup_dependence}(b1) - (b4). The peaks of
$v_y^{BN}/v_{yo}^E$ happen ahead of those of $J_a/J_o$ with increasing
$r_s$, which indicates that upstream particles subject to stronger
acceleration can detect downstream particles partially blocking the
hopper exit earlier. Thus, the corresponding peak of
$v_y^{BN}/v_{yo}^E$, signaling the downstream-upstream coupling, also
occurs earlier, as shown in Fig. \ref{fig:speedup_dependence}(b2).

\subsubsection{The geometry of the hopper exit}
\label{Effect3_waiting_room}
\begin{figure}
\includegraphics[width=0.40\textwidth]{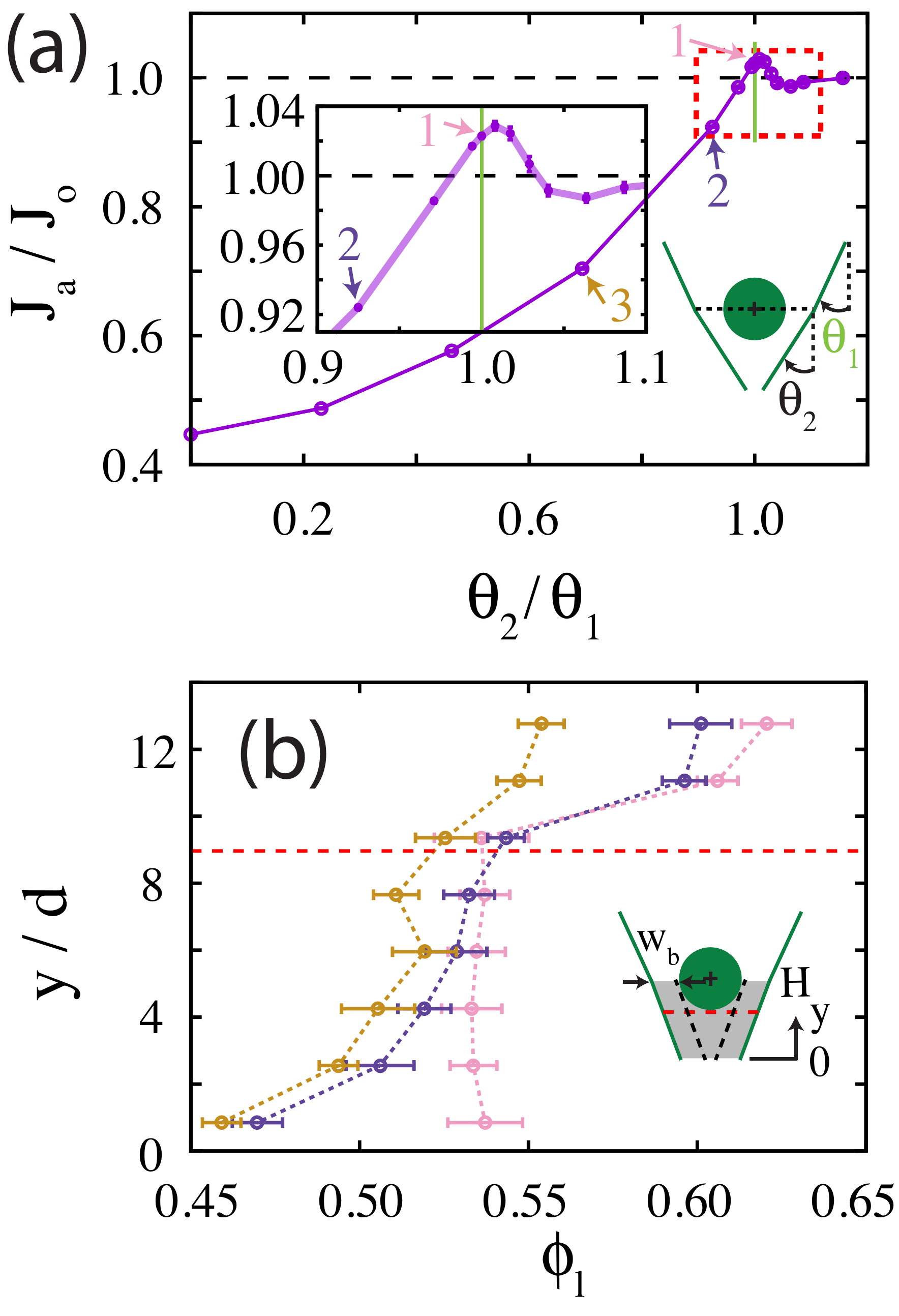}
\caption{\label{fig:kink_dependence_py0143} (Color online) (a)
  Normalized hopper flow rate $J_a/J_o$ measured at the exit of a
  hopper with its exit angle $\theta_2$ variable, containing a round
  obstacle placed at $H/d=13.612$, and $N=2048$. The driving strengths
  $\alpha_{x}=1.0$, $\alpha_y=0.333$, and the speed-up rate
  $r_s=1.0$. The green vertical line indicates where $\theta_2$ equals
  the hopper angle $\theta_1=0.4325$ and the two hopper walls are
  flat. The inset is a zoom-in of the red-dashed region. (b) Averaged
  local area packing fraction $\phi_l$ in the shaded zone in-between
  $y=[0,H]$ and next to the hopper wall within a stripe of width $w_b
  \approx 3.669d$, as shown in the inset. The three selected cases
  correspond to those identified in the inset of (a):
  $\theta_2/\theta_1=1.0$ (pink), $\approx 0.925$ (dark purple), and
  $\approx 0.694$ (brown). The horizontal dashed line indicates the
  lowest $y$ positions of the obstacle.}
\end{figure}

Lastly, we assess whether shrinking the geometry of the void space
from the obstacle to the hopper exit can effectively assist the
particle concentration mechanism and the flow rate enhancement. We
test this by varying the exit angle $\theta_2$ of a hopper with a
round obstacle placed at a fixed position where a flow rate peak
appears. Here, we chose the system with $N=2048$, $\alpha_x=1.0$,
$\alpha_y=0.333$, and $r_s=1.0$ so that $J_a/J_o$ peaks at
$H/d=13.612$, as previously shown by the black line in
Fig. \ref{fig:N_dependence_J}(a1). The results are shown in
Fig. \ref{fig:kink_dependence_py0143}(a). We find that the value of
the flow rate peak drops immediately as soon as the hopper walls
become uneven with an opening hopper exit, represented by
$\theta_2/\theta_1 < 1.0$, where $\theta_1$ is the fixed hopper
angle. These results prove that the shrinking geometry is crucial, a
conclusion that can also be inferred from the particle concentration
mechanism in Sec. \ref{peaking_mechanism}. In contrast, if $\theta_2$
is varied in the opposite direction to become uneven with a closing
hopper exit, represented by $\theta_2/\theta_1>1.0$, there exists a
very narrow range of $\theta_2/\theta_1$ between $1.0$ and $1.029$
that further enhances the peak of $J_a/J_o$ from the flat hopper
design. Beyond that, no enhanced flow rate, $J_a/J_o \ge 1$, is
observed due to the high clogging probability produced by the
increasingly narrow hopper exit. In
Fig. \ref{fig:kink_dependence_py0143}(b), we plot $\phi_l$, below the
obstacle and next to the hopper wall within a stripe of width $w_b$,
of three selected ${\theta _2}/{\theta _1} \le 1$ labeled in
Fig. \ref{fig:kink_dependence_py0143}(a). The results show that if
${\theta _2}/{\theta _1} < 1$, $\phi_l$ in this region cannot maintain
a constant value and decreases monotonically, again confirming the
necessity of the shrinking geometry of the hopper exit for creating a
flow rate peak.

\begin{figure}
\includegraphics[width=0.40\textwidth]{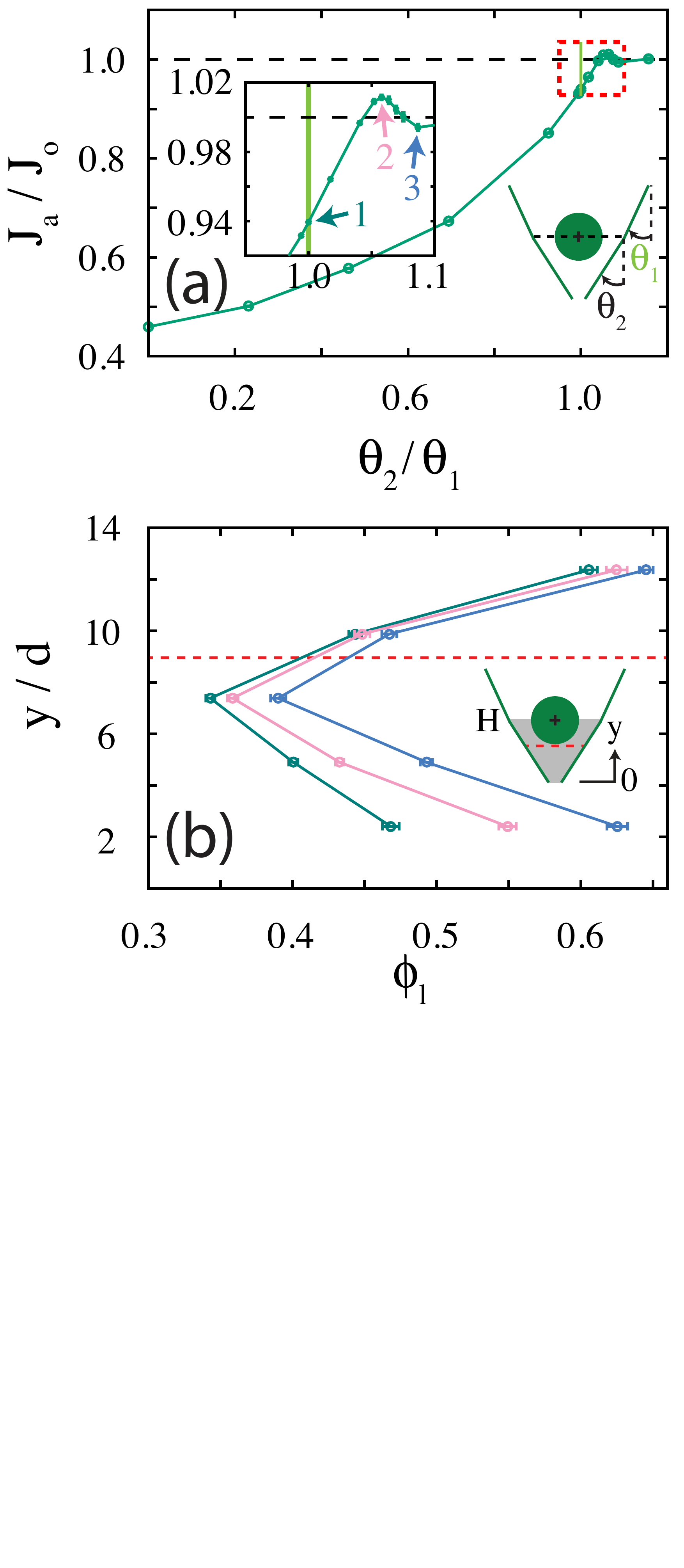}
\caption{\label{fig:kink_dependence_py0180} (Color online) (a) The
  same plot as Fig. \ref{fig:kink_dependence_py0143}, except
  $\alpha_y=0.439$. (b) Averaged local area packing fraction $\phi_l$
  in the shaded zone in-between $y=[0,H]$, as shown in the inset, of
  the three selected cases labeled in the inset of (a), where the
  value of $\theta_2$ is at 1 (dark green), 2 (pink), and 3
  (navy). The horizontal dashed line indicates the lowest y position
  of the obstacle.}
\end{figure}

In section \ref{Effect_merge_dependence}, we created an enhanced flow
rate peak in a system with $\alpha_y=0.439$ that originally exhibits
no such phenomenon using an anisotropic $\alpha_{x}$ below the
obstacle. Although effective, the strategy could be criticized for
being artificial and difficult to reproduce and verify
experimentally. Here we create a flow rate peak again in the same
system by narrowing the hopper below the obstacle, a setup which is
more practical. We show the enhanced $J_a/J_o$ in
Fig. \ref{fig:kink_dependence_py0180}(a), where the value of $J_a/J_o$
increases from $0.94$ to $1.01$, or by about $7\%$, when
$\theta_2/\theta_1$ increases from $1.0$ to $1.058$. It should be
noted that $J_o$ used for normalization decreases with increasing
$\theta_2$ because the hopper opening becomes narrower. Therefore,
this method is inherently at a disadvantage as it does not enhance the
unnormalized flow rate $J_a$. In
Fig. \ref{fig:kink_dependence_py0180}(b), we plot $\phi_l$ of three
selected cases with increasing $\theta_2$ before, at, and after the
peak of $J_a/J_o$, as labeled in
Fig. \ref{fig:kink_dependence_py0180}(a). Once again, we find a
familiar variation of $\phi_l$ at the hopper exit, similar to those
shown in Fig. \ref{fig:factorized_flowrate1}(b),
Fig. \ref{fig:factorized_flowrate2}(b2),
Fig. \ref{fig:awl_dependence_2}(a2, b1), and in our previous study
\cite{gao19}. The value of $\phi_l$ at the hopper exit when a peak of
$J_a/J_o$ occurs is between $0.5$ and $0.6$ in all studied cases.

\section{Conclusions}
\label{conclusions}
We investigate the phenomenon of the flow rate enhancement in a hopper
discharging athermal granular particles passing an obstacle placed
near its exit. Several competing mechanisms, such as interparticle
collaborative motion and particle acceleration due to gravity are
potentially responsible for the phenomenon. To decompose these
competing mechanisms, we leverage a probabilistic 2D Tetris-like
model, where particles move without creating overlaps between objects
in the system by following a position-update algorithm without
Newton's equations of motion. Our model preserves only the minimal
dynamics necessary to investigate particle motion within the free
space below the obstacle with or without acceleration due to an
external driving force, such as gravity.

The enhanced flow rate phenomenon still occurs in our probabilistic
model that switches off interparticle collaborative motion. We find
that the peaking phenomenon is limited to a system of at least a few
hundred of particles that pass through the channels between the
obstacle and the hopper walls, and then concentrate below the obstacle
on their way out of the hopper. Adjusting the height of the obstacle
in the hopper can reduce the amount of particles passing through the
channels between the obstacle and hopper walls and in turn decrease
the particle packing fraction at the hopper exit. Particles of lower
packing fraction exited the hopper faster. The flow rate enhancement
can be explained by the above mechanism if an obstacle is placed at an
optimal distance away from the hopper exit that mildly reduces the
particle packing fraction in exchange for a faster discharging
velocity. In all studied cases, the value of the local area packing
fraction $\phi_l$ at the hopper exit when a flow rate peak occurs is
within a range of $0.5$ and $0.6$, which corresponds to the
characteristic transition from $\phi_l^E < \phi_o^c$ to $\phi_l^E >
\phi_o^c$, where $\phi_o^c \approx 0.55$ is a characteristic area
packing fraction for the parameters chosen in this study. Too small
$\phi_l$ implies ineffective particle concentration mechanism, while
too large $\phi_l$ suggests interparticle repelling/clogging which are
both unfavorable to enhance the flow rate. The characteristic
transition suggests that particles unable to update their positions
form a cluster (two particles near to each other are treated as
clustered) that spans the hopper exit, as discussed in the appendix,
Sec. \ref{flow_property}.

Based on our findings, we then utilized the concentration mechanism by
artificially requiring particles below the obstacle to preferably move
away from the hopper walls in the horizontal direction. We proved that
this strategy can further amplify the value of the flow rate peak of
weakly-driven particles or even create one in a relatively
strongly-driven system that originally exhibits no peaking
phenomenon. This confirms that the particle concentration mechanism at
the hopper exit is indeed responsible for the flow rate
enhancement. Although artificial, it is possible to test the
anisotropic merging strategy experimentally using granular particles
guided by engraved guidelines on a 2D hopper surface or using metallic
particles responding to an external magnetic field
\cite{felipe17}. Adding a shortcut path through the obstacle which
allows particles to directly move towards the hopper exit interferes
with the particle concentration mechanism and generally degrades the
enhanced flow rate phenomenon.

Finally, we investigate the factors related to the waiting room effect
by decomposing it into three parts: the void space below the obstacle,
particle speed-up rate, and the exit geometry of the hopper. The
waiting room effect, wherein particles are slowed down by the obstacle
and then speed up within the void below it, is believed to contribute
to the higher flow rate observed in the peaking phenomenon. Our
results show that a flow rate peak still exists even when the defined
void space of the waiting room has been reduced to zero and, more
importantly, can be explained by the concentration
mechanism. Moreover, the enhanced flow rate can be augmented even
further if the channels between the obstacle and the hopper walls have
a shrinking geometry towards the hopper exit. There also exists a
narrow range of particle speed-up rate or hopper exit angles by which
an existing flow rate peak can be further improved. The latter can
also be utilized to create a flow rate peak manually.

All of these results support the conclusion that the concentration
mechanism at the hopper exit alone can create the hopper flow rate
enhancement, even if the waiting room effect is eliminated. However,
in an experimental setup containing frictional granular particles, the
concentration mechanism and the waiting room effect likely coexist,
and in tandem give rise to the flow rate peaking phenomenon.  We hope
the results of our Tetris-like model can motivate theoretical work. We
believe the particle concentration mechanism can be used for designing
hoppers that discharge granular particles more efficiently and expect
there to be broad industrial applications.

\section{Appendix: General flowing property of Tetris particles}
\label{flow_property}

\begin{figure}
\includegraphics[width=0.40\textwidth]{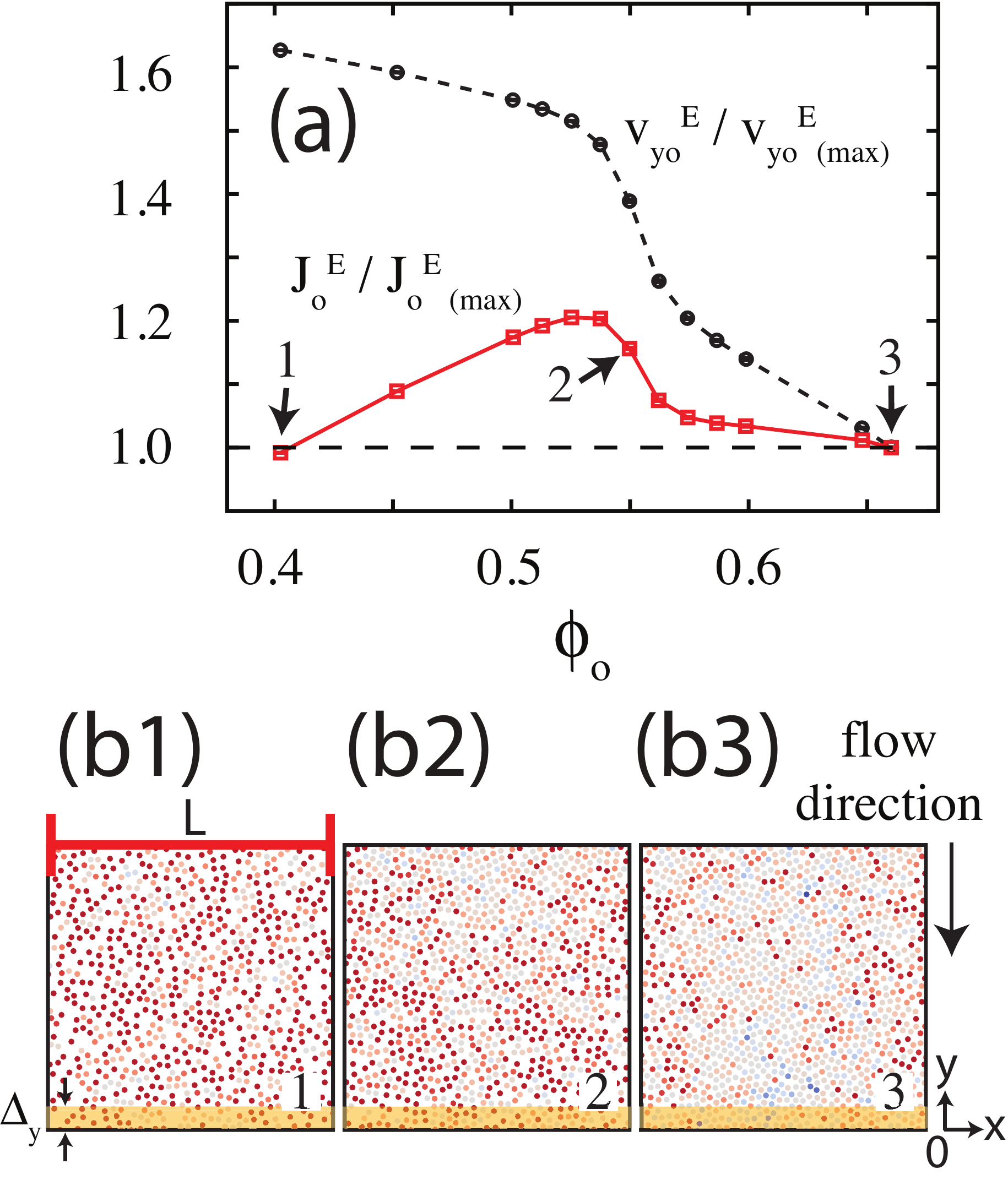}
\caption{\label{fig:exit_vel_calibration} (Color online) (a) Averaged
  particle velocity $v_{yo}^E$ (black circles), measured at $y_c =
  0.03125L = 1.25d$ and normalized by $v_{yo \ (max)}^E$ at $\phi_{o
    \ (max)} \approx 0.66$, in a square hopper of side $L$ and a
  hopper angle $\theta_1=0$. The calculated normalized flow rate (red
  squares) is obtained using ${J_o^E}/{J_{o \ (max)}^E} =
  (v_{yo}^E\phi _o)/(v_{yo \ (max)}^E\phi _{o \ (max)})$. A sudden
  slow-down of $v_{yo}^E$ which creates a local peak of ${J_o^E}$
  occurs when $\phi_o>\phi_o^c \approx 0.55$. Particles discharged
  from the hopper reenter it from its top boundary (red horizontal
  line in (b1)) with their x positions randomized. (b1-b3)
  Representative snapshots of the system as
  Fig. \ref{fig:factorized_flowrate1_snapshots}, except for the three
  cases indicated by arrows in (a), and $\Delta_y=0.0625L=2.5d$. Each
  mean and its error bar are obtained using $10$ trials.}
\end{figure}

In all studied cases, we observe that the range of local $\phi_l$ at
the hopper exit is between $0.5$ and $0.6$ when a peak of $J_a/J_o$
occurs. To understand if this specific range of $\phi_l$ is related to
a more intrinsic property of Tetris particles without the influence of
an obstacle or the shrinking hopper geometry, we chose a square hopper
containing no obstacle with a hopper angle $\theta_1=0$, equivalent to
a channel of constant width. In this setup, we can define a global
area packing fraction $\phi_o$ as the ratio of the total area of
particles in the hopper to its size. We measure the averaged particle
velocity $v_{yo}^E$ at the hopper exit and calculate the corresponding
flow rate $J_o^E=\phi _o v_{yo}^E w^E$, where $w^E=L$ is the constant
hopper width. Both quantities are normalized by their values at
$\phi_{o \ (max)} \approx 0.66$. The results are shown in
Fig. \ref{fig:exit_vel_calibration}.

In Fig. \ref{fig:exit_vel_calibration}(a), we can see that $v_{yo}^E$
exhibits a sudden slow-down when $\phi_o$ exceeds a characteristic
area packing fraction $\phi_o^c \approx 0.55$. Simultaneously, $J_o^E$
shows a peak right before $\phi_o>\phi_o^c$. In
Fig. \ref{fig:exit_vel_calibration}(b1) - (b3), we show the
representative snapshots of the three selected cases before, at, and
after $\phi_o^c$, labeled as 1, 2, and 3 in (a). Particles are colored
in blue by $n^f$ or in red by $n^s$ according to their consecutive
failed or successful position-update history. When $\phi_o<\phi_o^c$,
we can see that particles unable to update their positions (with a
nonzero $n^f$) form small and local clusters randomly distributed in
the hopper. On the other hand, when $\phi_o>\phi_o^c$, particles of
nonzero $n^f$ form a cluster across the system, especially along the
flow direction. The phenomenon of the sudden slow-down due to an
increasing $\phi_o$, and the ensuing frequent failures of particle
position-update and system size comparable clustering behavior seems
to be similar to the transition to slow dynamics where touching/nearly
touching particles form a network in a system governed by Newtonian
dynamics \cite{corey12,bassett18}. An investigation to fully quantify
this phenomenon of Tetris particles is left for future exploration.

\section{acknowledgments}
We thank Corey S. O'Hern for useful discussions. GJG acknowledges
financial support from National Taiwan University funding 104R7417,
MOST Grant No. 104-2218-E-002-019 (Taiwan), and computational facility
made available by the startup funding of Shizuoka University (Japan).

\bibliography{paper}

\providecommand{\noopsort}[1]{}\providecommand{\singleletter}[1]{#1}%
\begin{thebibliography}{24}%
\makeatletter
\providecommand \@ifxundefined [1]{%
 \@ifx{#1\undefined}
}%
\providecommand \@ifnum [1]{%
 \ifnum #1\expandafter \@firstoftwo
 \else \expandafter \@secondoftwo
 \fi
}%
\providecommand \@ifx [1]{%
 \ifx #1\expandafter \@firstoftwo
 \else \expandafter \@secondoftwo
 \fi
}%
\providecommand \natexlab [1]{#1}%
\providecommand \enquote  [1]{``#1''}%
\providecommand \bibnamefont  [1]{#1}%
\providecommand \bibfnamefont [1]{#1}%
\providecommand \citenamefont [1]{#1}%
\providecommand \href@noop [0]{\@secondoftwo}%
\providecommand \href [0]{\begingroup \@sanitize@url \@href}%
\providecommand \@href[1]{\@@startlink{#1}\@@href}%
\providecommand \@@href[1]{\endgroup#1\@@endlink}%
\providecommand \@sanitize@url [0]{\catcode `\\12\catcode `\$12\catcode
  `\&12\catcode `\#12\catcode `\^12\catcode `\_12\catcode `\%12\relax}%
\providecommand \@@startlink[1]{}%
\providecommand \@@endlink[0]{}%
\providecommand \url  [0]{\begingroup\@sanitize@url \@url }%
\providecommand \@url [1]{\endgroup\@href {#1}{\urlprefix }}%
\providecommand \urlprefix  [0]{URL }%
\providecommand \Eprint [0]{\href }%
\providecommand \doibase [0]{http://dx.doi.org/}%
\providecommand \selectlanguage [0]{\@gobble}%
\providecommand \bibinfo  [0]{\@secondoftwo}%
\providecommand \bibfield  [0]{\@secondoftwo}%
\providecommand \translation [1]{[#1]}%
\providecommand \BibitemOpen [0]{}%
\providecommand \bibitemStop [0]{}%
\providecommand \bibitemNoStop [0]{.\EOS\space}%
\providecommand \EOS [0]{\spacefactor3000\relax}%
\providecommand \BibitemShut  [1]{\csname bibitem#1\endcsname}%
\let\auto@bib@innerbib\@empty
\bibitem [{\citenamefont {Zuriguel}\ \emph {et~al.}(2011)\citenamefont
  {Zuriguel}, \citenamefont {Janda}, \citenamefont {Garcimart{\'i}n},
  \citenamefont {Lozano}, \citenamefont {Ar{\'e}valo},\ and\ \citenamefont
  {Maza}}]{zuriguel11}%
  \BibitemOpen
  \bibfield  {author} {\bibinfo {author} {\bibfnamefont {I.}~\bibnamefont
  {Zuriguel}}, \bibinfo {author} {\bibfnamefont {A.}~\bibnamefont {Janda}},
  \bibinfo {author} {\bibfnamefont {A.}~\bibnamefont {Garcimart{\'i}n}},
  \bibinfo {author} {\bibfnamefont {C.}~\bibnamefont {Lozano}}, \bibinfo
  {author} {\bibfnamefont {R.}~\bibnamefont {Ar{\'e}valo}}, \ and\ \bibinfo
  {author} {\bibfnamefont {D.}~\bibnamefont {Maza}},\ }\href@noop {} {\bibfield
   {journal} {\bibinfo  {journal} {Phys. Rev. Lett.}\ }\textbf {\bibinfo
  {volume} {107}},\ \bibinfo {pages} {278001} (\bibinfo {year}
  {2011})}\BibitemShut {NoStop}%
\bibitem [{\citenamefont {Zuriguel}\ \emph {et~al.}(2014)\citenamefont
  {Zuriguel}, \citenamefont {Parisi}, \citenamefont {Hidalgo}, \citenamefont
  {Lozano}, \citenamefont {Janda}, \citenamefont {Gago}, \citenamefont
  {Peralta}, \citenamefont {Ferrer}, \citenamefont {Pugnaloni}, \citenamefont
  {Cl{\'e}ment}, \citenamefont {Maza}, \citenamefont {Pagonabarraga},\ and\
  \citenamefont {Garcimartín}}]{zuriguel14}%
  \BibitemOpen
  \bibfield  {author} {\bibinfo {author} {\bibfnamefont {I.}~\bibnamefont
  {Zuriguel}}, \bibinfo {author} {\bibfnamefont {D.~R.}\ \bibnamefont
  {Parisi}}, \bibinfo {author} {\bibfnamefont {R.~C.}\ \bibnamefont {Hidalgo}},
  \bibinfo {author} {\bibfnamefont {C.}~\bibnamefont {Lozano}}, \bibinfo
  {author} {\bibfnamefont {A.}~\bibnamefont {Janda}}, \bibinfo {author}
  {\bibfnamefont {P.~A.}\ \bibnamefont {Gago}}, \bibinfo {author}
  {\bibfnamefont {J.~P.}\ \bibnamefont {Peralta}}, \bibinfo {author}
  {\bibfnamefont {L.~M.}\ \bibnamefont {Ferrer}}, \bibinfo {author}
  {\bibfnamefont {L.~A.}\ \bibnamefont {Pugnaloni}}, \bibinfo {author}
  {\bibfnamefont {E.}~\bibnamefont {Cl{\'e}ment}}, \bibinfo {author}
  {\bibfnamefont {D.}~\bibnamefont {Maza}}, \bibinfo {author} {\bibfnamefont
  {I.}~\bibnamefont {Pagonabarraga}}, \ and\ \bibinfo {author} {\bibfnamefont
  {A.}~\bibnamefont {Garcimartín}},\ }\href@noop {} {\bibfield  {journal}
  {\bibinfo  {journal} {Sci. Rep.}\ }\textbf {\bibinfo {volume} {4}},\ \bibinfo
  {pages} {7324} (\bibinfo {year} {2014})}\BibitemShut {NoStop}%
\bibitem [{\citenamefont {Lozano}\ \emph {et~al.}(2012)\citenamefont {Lozano},
  \citenamefont {Janda}, \citenamefont {Garcimart{\'i}n}, \citenamefont
  {Maza},\ and\ \citenamefont {Zuriguel}}]{lozano12}%
  \BibitemOpen
  \bibfield  {author} {\bibinfo {author} {\bibfnamefont {C.}~\bibnamefont
  {Lozano}}, \bibinfo {author} {\bibfnamefont {A.}~\bibnamefont {Janda}},
  \bibinfo {author} {\bibfnamefont {A.}~\bibnamefont {Garcimart{\'i}n}},
  \bibinfo {author} {\bibfnamefont {D.}~\bibnamefont {Maza}}, \ and\ \bibinfo
  {author} {\bibfnamefont {I.}~\bibnamefont {Zuriguel}},\ }\href@noop {}
  {\bibfield  {journal} {\bibinfo  {journal} {Phys. Rev. E}\ }\textbf {\bibinfo
  {volume} {86}},\ \bibinfo {pages} {031306} (\bibinfo {year}
  {2012})}\BibitemShut {NoStop}%
\bibitem [{\citenamefont {Alonso-Marroquin}\ \emph {et~al.}(2012)\citenamefont
  {Alonso-Marroquin}, \citenamefont {Azeezullah}, \citenamefont
  {Galindo-Torres},\ and\ \citenamefont {Olsen-Kettle}}]{alonso-marroquin12}%
  \BibitemOpen
  \bibfield  {author} {\bibinfo {author} {\bibfnamefont {F.}~\bibnamefont
  {Alonso-Marroquin}}, \bibinfo {author} {\bibfnamefont {S.~I.}\ \bibnamefont
  {Azeezullah}}, \bibinfo {author} {\bibfnamefont {S.~A.}\ \bibnamefont
  {Galindo-Torres}}, \ and\ \bibinfo {author} {\bibfnamefont {L.~M.}\
  \bibnamefont {Olsen-Kettle}},\ }\href@noop {} {\bibfield  {journal} {\bibinfo
   {journal} {Phys. Rev. E}\ }\textbf {\bibinfo {volume} {85}},\ \bibinfo
  {pages} {020301} (\bibinfo {year} {2012})}\BibitemShut {NoStop}%
\bibitem [{\citenamefont {Pastor}\ \emph {et~al.}(2015)\citenamefont {Pastor},
  \citenamefont {Garcimart{\'i}n}, \citenamefont {Gago}, \citenamefont {amd
  C{\'e}sar Mart{\'i}n-G{\'o}mez}, \citenamefont {Ferrer}, \citenamefont
  {Maza}, \citenamefont {Parisi}, \citenamefont {Pugnaloni},\ and\
  \citenamefont {Zuriguel}}]{zuriguel15}%
  \BibitemOpen
  \bibfield  {author} {\bibinfo {author} {\bibfnamefont {J.~M.}\ \bibnamefont
  {Pastor}}, \bibinfo {author} {\bibfnamefont {A.}~\bibnamefont
  {Garcimart{\'i}n}}, \bibinfo {author} {\bibfnamefont {P.~A.}\ \bibnamefont
  {Gago}}, \bibinfo {author} {\bibfnamefont {J.~P.~P.}\ \bibnamefont {amd
  C{\'e}sar Mart{\'i}n-G{\'o}mez}}, \bibinfo {author} {\bibfnamefont {L.~M.}\
  \bibnamefont {Ferrer}}, \bibinfo {author} {\bibfnamefont {D.}~\bibnamefont
  {Maza}}, \bibinfo {author} {\bibfnamefont {D.~R.}\ \bibnamefont {Parisi}},
  \bibinfo {author} {\bibfnamefont {L.~A.}\ \bibnamefont {Pugnaloni}}, \ and\
  \bibinfo {author} {\bibfnamefont {I.}~\bibnamefont {Zuriguel}},\ }\href@noop
  {} {\bibfield  {journal} {\bibinfo  {journal} {Phys. Rev. E}\ }\textbf
  {\bibinfo {volume} {92}},\ \bibinfo {pages} {062817} (\bibinfo {year}
  {2015})}\BibitemShut {NoStop}%
\bibitem [{\citenamefont {Murray}\ and\ \citenamefont
  {Alonso-Marroquin}(2016)}]{alonso-marroquin16}%
  \BibitemOpen
  \bibfield  {author} {\bibinfo {author} {\bibfnamefont {A.}~\bibnamefont
  {Murray}}\ and\ \bibinfo {author} {\bibfnamefont {F.}~\bibnamefont
  {Alonso-Marroquin}},\ }\href@noop {} {\bibfield  {journal} {\bibinfo
  {journal} {Paper in Physics}\ }\textbf {\bibinfo {volume} {8}},\ \bibinfo
  {pages} {080003} (\bibinfo {year} {2016})}\BibitemShut {NoStop}%
\bibitem [{\citenamefont {Helbing}\ \emph {et~al.}(2000)\citenamefont
  {Helbing}, \citenamefont {Farkas},\ and\ \citenamefont {Vicsek}}]{helbing00}%
  \BibitemOpen
  \bibfield  {author} {\bibinfo {author} {\bibfnamefont {D.}~\bibnamefont
  {Helbing}}, \bibinfo {author} {\bibfnamefont {I.}~\bibnamefont {Farkas}}, \
  and\ \bibinfo {author} {\bibfnamefont {T.}~\bibnamefont {Vicsek}},\
  }\href@noop {} {\bibfield  {journal} {\bibinfo  {journal} {Nature}\ }\textbf
  {\bibinfo {volume} {407}},\ \bibinfo {pages} {487} (\bibinfo {year}
  {2000})}\BibitemShut {NoStop}%
\bibitem [{\citenamefont {Escobar}\ and\ \citenamefont {Rosa}(2003)}]{rosa03}%
  \BibitemOpen
  \bibfield  {author} {\bibinfo {author} {\bibfnamefont {R.}~\bibnamefont
  {Escobar}}\ and\ \bibinfo {author} {\bibfnamefont {A.~D.~L.}\ \bibnamefont
  {Rosa}},\ }\href@noop {} {\emph {\bibinfo {title} {Advances in Artificial
  Life}}}\ (\bibinfo  {publisher} {Springer, Berlin},\ \bibinfo {year} {2003})\
  pp.\ \bibinfo {pages} {97--106}\BibitemShut {NoStop}%
\bibitem [{\citenamefont {Helbing}\ \emph {et~al.}(2005)\citenamefont
  {Helbing}, \citenamefont {Buzna}, \citenamefont {Johansson},\ and\
  \citenamefont {Werner}}]{helbing05}%
  \BibitemOpen
  \bibfield  {author} {\bibinfo {author} {\bibfnamefont {D.}~\bibnamefont
  {Helbing}}, \bibinfo {author} {\bibfnamefont {L.}~\bibnamefont {Buzna}},
  \bibinfo {author} {\bibfnamefont {A.}~\bibnamefont {Johansson}}, \ and\
  \bibinfo {author} {\bibfnamefont {T.}~\bibnamefont {Werner}},\ }\href@noop {}
  {\bibfield  {journal} {\bibinfo  {journal} {Transp. Sci.}\ }\textbf {\bibinfo
  {volume} {39}},\ \bibinfo {pages} {1} (\bibinfo {year} {2005})}\BibitemShut
  {NoStop}%
\bibitem [{\citenamefont {Garcimart{\'i}n}\ \emph {et~al.}(2015)\citenamefont
  {Garcimart{\'i}n}, \citenamefont {Pastor}, \citenamefont {Ferrer},
  \citenamefont {Ramos}, \citenamefont {Mart{\'i}n-G{\'o}mez},\ and\
  \citenamefont {Zuriguel}}]{zuriguel15_1}%
  \BibitemOpen
  \bibfield  {author} {\bibinfo {author} {\bibfnamefont {A.}~\bibnamefont
  {Garcimart{\'i}n}}, \bibinfo {author} {\bibfnamefont {J.~M.}\ \bibnamefont
  {Pastor}}, \bibinfo {author} {\bibfnamefont {L.~M.}\ \bibnamefont {Ferrer}},
  \bibinfo {author} {\bibfnamefont {J.~J.}\ \bibnamefont {Ramos}}, \bibinfo
  {author} {\bibfnamefont {C.}~\bibnamefont {Mart{\'i}n-G{\'o}mez}}, \ and\
  \bibinfo {author} {\bibfnamefont {I.}~\bibnamefont {Zuriguel}},\ }\href@noop
  {} {\bibfield  {journal} {\bibinfo  {journal} {Phys. Rev. E}\ }\textbf
  {\bibinfo {volume} {91}},\ \bibinfo {pages} {022808} (\bibinfo {year}
  {2015})}\BibitemShut {NoStop}%
\bibitem [{\citenamefont {Zuriguel}\ \emph {et~al.}(2016)\citenamefont
  {Zuriguel}, \citenamefont {Olivares}, \citenamefont {Pastor}, \citenamefont
  {Mart{\'i}n-G{\'o}mez}, \citenamefont {Ferrer}, \citenamefont {Ramos},\ and\
  \citenamefont {Garcimart{\'i}n}}]{zuriguel16}%
  \BibitemOpen
  \bibfield  {author} {\bibinfo {author} {\bibfnamefont {I.}~\bibnamefont
  {Zuriguel}}, \bibinfo {author} {\bibfnamefont {J.}~\bibnamefont {Olivares}},
  \bibinfo {author} {\bibfnamefont {J.~M.}\ \bibnamefont {Pastor}}, \bibinfo
  {author} {\bibfnamefont {C.}~\bibnamefont {Mart{\'i}n-G{\'o}mez}}, \bibinfo
  {author} {\bibfnamefont {L.~M.}\ \bibnamefont {Ferrer}}, \bibinfo {author}
  {\bibfnamefont {J.~J.}\ \bibnamefont {Ramos}}, \ and\ \bibinfo {author}
  {\bibfnamefont {A.}~\bibnamefont {Garcimart{\'i}n}},\ }\href@noop {}
  {\bibfield  {journal} {\bibinfo  {journal} {Phys. Rev. E}\ }\textbf {\bibinfo
  {volume} {94}},\ \bibinfo {pages} {032302} (\bibinfo {year}
  {2016})}\BibitemShut {NoStop}%
\bibitem [{\citenamefont {Endo}\ and\ \citenamefont
  {Katsuragi}(2017)}]{katsuragi17}%
  \BibitemOpen
  \bibfield  {author} {\bibinfo {author} {\bibfnamefont {K.}~\bibnamefont
  {Endo}}\ and\ \bibinfo {author} {\bibfnamefont {H.}~\bibnamefont
  {Katsuragi}},\ }\href@noop {} {\bibfield  {journal} {\bibinfo  {journal} {EPJ
  Web Conf.}\ }\textbf {\bibinfo {volume} {140}},\ \bibinfo {pages} {03004}
  (\bibinfo {year} {2017})}\BibitemShut {NoStop}%
\bibitem [{\citenamefont {Endo}\ \emph {et~al.}(2018)\citenamefont {Endo},
  \citenamefont {Reddy},\ and\ \citenamefont {Katsuragi}}]{katsuragi18}%
  \BibitemOpen
  \bibfield  {author} {\bibinfo {author} {\bibfnamefont {K.}~\bibnamefont
  {Endo}}, \bibinfo {author} {\bibfnamefont {K.~A.}\ \bibnamefont {Reddy}}, \
  and\ \bibinfo {author} {\bibfnamefont {H.}~\bibnamefont {Katsuragi}},\
  }\href@noop {} {\bibfield  {journal} {\bibinfo  {journal} {Phys. Rev.
  Fluids}\ }\textbf {\bibinfo {volume} {2}},\ \bibinfo {pages} {094302}
  (\bibinfo {year} {2018})}\BibitemShut {NoStop}%
\bibitem [{\citenamefont {Gao}\ \emph {et~al.}(2019)\citenamefont {Gao},
  \citenamefont {Blawzdziewicz}, \citenamefont {Holcomb},\ and\ \citenamefont
  {Ogata}}]{gao19}%
  \BibitemOpen
  \bibfield  {author} {\bibinfo {author} {\bibfnamefont {G.~J.}\ \bibnamefont
  {Gao}}, \bibinfo {author} {\bibfnamefont {J.}~\bibnamefont {Blawzdziewicz}},
  \bibinfo {author} {\bibfnamefont {M.~C.}\ \bibnamefont {Holcomb}}, \ and\
  \bibinfo {author} {\bibfnamefont {S.}~\bibnamefont {Ogata}},\ }\href@noop {}
  {\bibfield  {journal} {\bibinfo  {journal} {Granul. Matter.}\ }\textbf
  {\bibinfo {volume} {21: 25}} (\bibinfo {year} {2019})}\BibitemShut {NoStop}%
\bibitem [{\citenamefont {Gao}(2018)}]{gao18}%
  \BibitemOpen
  \bibfield  {author} {\bibinfo {author} {\bibfnamefont {G.~J.}\ \bibnamefont
  {Gao}},\ }\href@noop {} {\bibfield  {journal} {\bibinfo  {journal} {J. Phys.
  Soc. Jpn.}\ }\textbf {\bibinfo {volume} {87}},\ \bibinfo {pages} {114401}
  (\bibinfo {year} {2018})}\BibitemShut {NoStop}%
\bibitem [{\citenamefont {Caglioti}\ \emph {et~al.}(1997)\citenamefont
  {Caglioti}, \citenamefont {Loreto}, \citenamefont {Herrmann},\ and\
  \citenamefont {Nicodemi}}]{nicodemi97}%
  \BibitemOpen
  \bibfield  {author} {\bibinfo {author} {\bibfnamefont {E.}~\bibnamefont
  {Caglioti}}, \bibinfo {author} {\bibfnamefont {V.}~\bibnamefont {Loreto}},
  \bibinfo {author} {\bibfnamefont {H.~J.}\ \bibnamefont {Herrmann}}, \ and\
  \bibinfo {author} {\bibfnamefont {M.}~\bibnamefont {Nicodemi}},\ }\href@noop
  {} {\bibfield  {journal} {\bibinfo  {journal} {Phys. Rev. Lett.}\ }\textbf
  {\bibinfo {volume} {79}},\ \bibinfo {pages} {1575} (\bibinfo {year}
  {1997})}\BibitemShut {NoStop}%
\bibitem [{\citenamefont {Clauset}\ \emph {et~al.}(2009)\citenamefont
  {Clauset}, \citenamefont {Shalizi},\ and\ \citenamefont {Newman}}]{newman09}%
  \BibitemOpen
  \bibfield  {author} {\bibinfo {author} {\bibfnamefont {A.}~\bibnamefont
  {Clauset}}, \bibinfo {author} {\bibfnamefont {C.~R.}\ \bibnamefont
  {Shalizi}}, \ and\ \bibinfo {author} {\bibfnamefont {M.~E.~J.}\ \bibnamefont
  {Newman}},\ }\href@noop {} {\bibfield  {journal} {\bibinfo  {journal} {SIAM
  Review}\ }\textbf {\bibinfo {volume} {51}},\ \bibinfo {pages} {661} (\bibinfo
  {year} {2009})}\BibitemShut {NoStop}%
\bibitem [{\citenamefont {Gardel}\ \emph {et~al.}(2009)\citenamefont {Gardel},
  \citenamefont {Seitaridou}, \citenamefont {Facto}, \citenamefont {Keene},
  \citenamefont {Hattam}, \citenamefont {Easwar},\ and\ \citenamefont
  {MENON}}]{menon09}%
  \BibitemOpen
  \bibfield  {author} {\bibinfo {author} {\bibfnamefont {E.}~\bibnamefont
  {Gardel}}, \bibinfo {author} {\bibfnamefont {E.}~\bibnamefont {Seitaridou}},
  \bibinfo {author} {\bibfnamefont {K.}~\bibnamefont {Facto}}, \bibinfo
  {author} {\bibfnamefont {E.}~\bibnamefont {Keene}}, \bibinfo {author}
  {\bibfnamefont {K.}~\bibnamefont {Hattam}}, \bibinfo {author} {\bibfnamefont
  {N.}~\bibnamefont {Easwar}}, \ and\ \bibinfo {author} {\bibfnamefont
  {N.}~\bibnamefont {MENON}},\ }\href@noop {} {\bibfield  {journal} {\bibinfo
  {journal} {Phil. Trans. R. Soc. A}\ }\textbf {\bibinfo {volume} {367}},\
  \bibinfo {pages} {5109} (\bibinfo {year} {2009})}\BibitemShut {NoStop}%
\bibitem [{\citenamefont {Tewari}\ \emph {et~al.}(2013)\citenamefont {Tewari},
  \citenamefont {Dichter},\ and\ \citenamefont {Chakraborty}}]{tewari13}%
  \BibitemOpen
  \bibfield  {author} {\bibinfo {author} {\bibfnamefont {S.}~\bibnamefont
  {Tewari}}, \bibinfo {author} {\bibfnamefont {M.}~\bibnamefont {Dichter}}, \
  and\ \bibinfo {author} {\bibfnamefont {B.}~\bibnamefont {Chakraborty}},\
  }\href@noop {} {\bibfield  {journal} {\bibinfo  {journal} {Soft Matter}\
  }\textbf {\bibinfo {volume} {9}},\ \bibinfo {pages} {5016} (\bibinfo {year}
  {2013})}\BibitemShut {NoStop}%
\bibitem [{\citenamefont {Thomas}\ and\ \citenamefont
  {Durian}(2016)}]{durian16}%
  \BibitemOpen
  \bibfield  {author} {\bibinfo {author} {\bibfnamefont {C.~C.}\ \bibnamefont
  {Thomas}}\ and\ \bibinfo {author} {\bibfnamefont {D.~J.}\ \bibnamefont
  {Durian}},\ }\href@noop {} {\bibfield  {journal} {\bibinfo  {journal} {Phys.
  Rev. E}\ }\textbf {\bibinfo {volume} {94}},\ \bibinfo {pages} {022901}
  (\bibinfo {year} {2016})}\BibitemShut {NoStop}%
\bibitem [{\citenamefont {Zuriguel}\ \emph {et~al.}(2019)\citenamefont
  {Zuriguel}, \citenamefont {Maza}, \citenamefont {Janda}, \citenamefont
  {Hidalgo},\ and\ \citenamefont {Garcimart{\'i}n}}]{zuriguel19}%
  \BibitemOpen
  \bibfield  {author} {\bibinfo {author} {\bibfnamefont {I.}~\bibnamefont
  {Zuriguel}}, \bibinfo {author} {\bibfnamefont {D.}~\bibnamefont {Maza}},
  \bibinfo {author} {\bibfnamefont {A.}~\bibnamefont {Janda}}, \bibinfo
  {author} {\bibfnamefont {R.~C.}\ \bibnamefont {Hidalgo}}, \ and\ \bibinfo
  {author} {\bibfnamefont {A.}~\bibnamefont {Garcimart{\'i}n}},\ }\href@noop {}
  {\bibfield  {journal} {\bibinfo  {journal} {Granul. Matter.}\ }\textbf
  {\bibinfo {volume} {21: 47}} (\bibinfo {year} {2019})}\BibitemShut {NoStop}%
\bibitem [{\citenamefont {Hern{\'a}ndez-Enríquez}\ \emph
  {et~al.}(2017)\citenamefont {Hern{\'a}ndez-Enríquez}, \citenamefont
  {Lumay},\ and\ \citenamefont {Pacheco-V{\'a}zquez}}]{felipe17}%
  \BibitemOpen
  \bibfield  {author} {\bibinfo {author} {\bibfnamefont {D.}~\bibnamefont
  {Hern{\'a}ndez-Enríquez}}, \bibinfo {author} {\bibfnamefont
  {G.}~\bibnamefont {Lumay}}, \ and\ \bibinfo {author} {\bibfnamefont
  {F.}~\bibnamefont {Pacheco-V{\'a}zquez}},\ }\href@noop {} {\bibfield
  {journal} {\bibinfo  {journal} {EPJ Web Conf.}\ }\textbf {\bibinfo {volume}
  {140}},\ \bibinfo {pages} {03089} (\bibinfo {year} {2017})}\BibitemShut
  {NoStop}%
\bibitem [{\citenamefont {Shen}\ \emph {et~al.}(2012)\citenamefont {Shen},
  \citenamefont {Schreck}, \citenamefont {Chakraborty}, \citenamefont {Freed},\
  and\ \citenamefont {O'Hern}}]{corey12}%
  \BibitemOpen
  \bibfield  {author} {\bibinfo {author} {\bibfnamefont {T.}~\bibnamefont
  {Shen}}, \bibinfo {author} {\bibfnamefont {C.}~\bibnamefont {Schreck}},
  \bibinfo {author} {\bibfnamefont {B.}~\bibnamefont {Chakraborty}}, \bibinfo
  {author} {\bibfnamefont {D.~E.}\ \bibnamefont {Freed}}, \ and\ \bibinfo
  {author} {\bibfnamefont {C.~S.}\ \bibnamefont {O'Hern}},\ }\href@noop {}
  {\bibfield  {journal} {\bibinfo  {journal} {Phys. Rev. E}\ }\textbf {\bibinfo
  {volume} {86}},\ \bibinfo {pages} {041303} (\bibinfo {year}
  {2012})}\BibitemShut {NoStop}%
\bibitem [{\citenamefont {Papadopoulos}\ \emph {et~al.}(2018)\citenamefont
  {Papadopoulos}, \citenamefont {Porter}, \citenamefont {Daniels},\ and\
  \citenamefont {Bassett}}]{bassett18}%
  \BibitemOpen
  \bibfield  {author} {\bibinfo {author} {\bibfnamefont {L.}~\bibnamefont
  {Papadopoulos}}, \bibinfo {author} {\bibfnamefont {M.~A.}\ \bibnamefont
  {Porter}}, \bibinfo {author} {\bibfnamefont {K.~E.}\ \bibnamefont {Daniels}},
  \ and\ \bibinfo {author} {\bibfnamefont {D.~S.}\ \bibnamefont {Bassett}},\
  }\href@noop {} {\bibfield  {journal} {\bibinfo  {journal} {J. Complex Netw.}\
  }\textbf {\bibinfo {volume} {6}},\ \bibinfo {pages} {485} (\bibinfo {year}
  {2018})}\BibitemShut {NoStop}%
\end{thebibliography}%

\end{document}